\documentclass{article}

\usepackage{bm}
\usepackage{amssymb}
\usepackage{amsthm}
\usepackage{longtable}
\usepackage{amsfonts}
\usepackage{geometry}
\usepackage{multirow}
\usepackage{multicol}
\usepackage{graphicx}
\usepackage{epsfig}
\usepackage{epstopdf}
\usepackage{subfigure}
\usepackage{color}
\usepackage{ulem}
\usepackage{booktabs}
\usepackage{rotating}
\usepackage[toc,page]{appendix}
\usepackage{cite}
\renewcommand{\theequation}{\arabic{section}.\arabic{equation}}
\renewcommand{\theequation}{\arabic{section}.\arabic{equation}}

\def\b{\beta}

\def\e{\epsilon}

\def\l{\lambda}

\def\s{\sigma}

\def\V={{{\bf\rm{V}}}}

\def\Cb{\mathbb{C}}

\def\beq{\begin{equation}}
\def\eeq{\end{equation}}
\def\bea{\begin{eqnarray}}
\def\eea{\end{eqnarray}}
\def\ba{\begin{array}}
\def\ea{\end{array}}
\def\no{\nonumber}

\def\lt{\left}
\def\rt{\right}

\title{{ \bf $T-W$ relation and free energy of the Heisenberg chain at a finite temperature}}
\author{}
\begin{document}

\date{}
\maketitle
\begin{center}
{\large\bf
Pengcheng Lu${}^{a,b}$, Yi Qiao${}^{b}$, Junpeng Cao${}^{b,c,d,e}$,  Wen-Li
Yang${}^{a,e,f}\footnote{Corresponding author:
wlyang@nwu.edu.cn}$, Kangjie Shi${}^{a}$ and Yupeng
Wang${}^{b,c,e,g}\footnote{Corresponding author: yupeng@iphy.ac.cn}$}
\\
\vspace{0.46truecm}

${}^a$ Institute of Modern Physics, Northwest University,
     Xian 710127, China\\

${}^b$ Beijing National Laboratory for Condensed Matter
           Physics, Institute of Physics, Chinese Academy of Sciences, Beijing
           100190, China\\
     ${}^c$ School of Physical Sciences, University of Chinese Academy of
Sciences, Beijing, China\\
     ${}^d$ Songshan Lake Materials Laboratory, Dongguan, Guangdong 523808, China \\
     ${}^e$ Peng Huanwu Center for Fundamental Theory, Xian 710127, China\\
     ${}^f$ Shaanxi Key Laboratory for Theoretical Physics Frontiers,  Xian 710127, China\\
     ${}^g$ The Yangtze River Delta Physics Research Center, Liyang, Jiangsu, China
\end{center}

\vspace{0.46truecm}

\begin{abstract}
A new nonlinear integral equation (NLIE) describing the thermodynamics of the Heisenberg spin chain is derived based on the $t-W$ relation
of the quantum transfer matrices. The free energy of the system in a magnetic field is thus obtained by solving the NLIE. This method can be generalized to other lattice quantum integrable models. Taking the $SU(3)$-invariant quantum spin chain  as an example, we construct the corresponding NLIEs and compute the free energy. The present results coincide exactly with those obtained via other methods previously.

\vspace{1truecm} \noindent {\it PACS:}
75.10.Pq, 03.65.Vf, 71.10.Pm

\end{abstract}
\newpage
\section{Introduction}
\label{intro} \setcounter{equation}{0}

Quantum integrable systems (or exactly solvable models \cite{Bax82}) play important roles in investigating  some non-pertubative properties of quantum
field/string theory such as  the planar ${\cal{N}}=4$ super-symmetric Yang-Mills (SYM) theory and the planar
AdS/CFT \cite{Mal98, Bei12} (see also references therein). They also enhance our understanding of quantum phase transitions and critical phenomena
in statistical physics \cite{Gie05,Sir09}, condensed matter physics \cite{Duk04} and  cold atom systems \cite{Gua13}. In the past decades, several theoretical methods \cite{Bet31,Bax82,Skl80,kor97,Res83,Skl89,Skl95,Wan15,Bas13,Ava15}  have been  developed to approach  eigenvalue problem of quantum integrable models. A method to approach thermodynamic properties of quantum integrable models was first achieved by Yang and Yang for the quantum Bose gas \cite{YY.1969,PhysRevA.2.154} based on the Bethe ansatz solution \cite{lieb1,lieb2}. Later, the method (now known as thermodynamic Bethe ansatz (TBA)) was  extended by Gaudin\,\cite{PhysRevLett.26.1301} and Takahashi\,\cite{PTP.46.401,takahashi2005thermodynamics}  to investigate the thermodynamics of the Heisenberg spin chain. With their methods, the free energy was finally found to be encoded by a set of infinitely many nonlinear integral equations (NLIEs). The numerical studies of these equations need some kind of truncation scheme \cite{PhysLettA.1986133,PhysRevB.45.5293,KPS1994,LEE1994112}. An alternative approach, the so-called quantum transfer matrix (QTM) method  \cite{PhysRevB.31.2957,Kom87,Suzuki90,klumper1992free,klumper1998free,PhysRevLett.69.2313,Ess05} has also been proposed. In the QTM formalism, a one-dimensional quantum system at a finite temperature can be mapped into a classical system on two-dimensional inhomogeneous lattice by the  Trotter-Suzuki mapping \cite{PhysRevB.31.2957}. The free energy of the quantum system can be expressed by the largest eigenvalue of the quantum transfer matrix and the next-largest eigenvalue provides the correlation length \cite{Ess05}. The advantage of QTM is that only a finite number of NLIEs are needed. Indeed, with the fusion hierarchy \cite{PhysRevB.31.2957,klumper1992free,klumper1998free,Kirillov.1987,Bazhanov.1990,KLUMPER1992304,kuniba1994functional,Tsuboi.1997,TSUBOI1998565,JUTTNER1998581}
the TBA equations can be rederived. Moreover, the QTM method also allows ones to study the finite-size corrections \cite{Klumper.1990,Klumper.1991,Klumper.1993SIX,Benz.J2005,Suzuki.1999,G.Juttner1997,PhysRevLett.81.4975,Klumper.1997PS,RIBEIRO2008,SUZUKI1998,FUJII1999}. Similar NLIE \cite{takahashi2001physics,Tomotoshi.Nishino1995,Tsuboi.2003,PhysRevLett.89.117201} exists to obtain the high-temperature expansions of the free energy  up to a very high order. In addition, the transfer-matrix renormalization-group \cite{Tomotoshi.Nishino1995} (TMRG) has been shown to be a very powerful numerical method to study the thermodynamics of various quantum spin chain systems\,\cite{PhysRevB.86.235102,PhysRevB.56.5061,Bursill.1996,PhysRevB.58.9142,PhysRevB.60.359,PhysRevB.69.104428,PhysRevB.74.134425,Sota.2010}.

Recently, a novel $t-W$ method has been proposed to calculate physical properties of quantum integrable systems with or without $U(1)$ symmetry \cite{PhysRevB.102.085115,Qia21}. The key point of this method  lies in that a single $t-W$ relation determines the whole spectrum of the transfer matrix and the roots possess well-defined patterns. In this paper, we will construct the $t-W$ relation of the quantum transfer matrix. By analysing the root patterns of the quantum transfer matrix, a new NLIE describing the thermodynamics can be derived straightforwardly based on the $t-W$ relation.

Let us consider the Hamiltonian of the periodic Heisenberg spin chain in anti-ferromagnetic regime ($J>0$):
\begin{eqnarray}
H = J\sum_{n=1}^L
  (\sigma_n^x\sigma_{n+1}^x+\sigma_n^y\sigma_{n+1}^y
  +\sigma_n^z\sigma_{n+1}^z)+\frac{h}{2}\sum_{j=1}^L\sigma_j^z, \label{xxzh}
\end{eqnarray}
where
\bea
\sigma^{\alpha}_{1+L}= \sigma^{\alpha}_{1},\quad {\rm  for}\quad \alpha=x,y,z,\label{BC}
\eea
and $\sigma^x,\,\sigma^y,\,\sigma^z$ are Pauli matrices.  The model is one of the best studied paradigmatic models in quantum integrable systems and still remains a source of inspiration and fascinating new progress of quantum integrable systems.

The paper is organized as follows. Section 2 serves as an introduction of our notations and some basic ingredients. We also briefly review that the partition function of the Heisenberg chain at a finite temperature is expressed
in the QTM formalism. In Section 3,  we derive the $t-W$ relation of the transfer matrix via the fusion technique. With the help of the resulting $t-W$  relation and some asymptotical behaviors of eigenvalues of
the transfer matrices, we obtain the Bethe-ansatz-like equations (BAEs), which may completely determine eigenvalues. In Section 4, based on the root distributions  of eigenvalues corresponding  to the state with the maximus $|\Lambda^{(Q)}(0)|$, we derive a new nonlinear integral equation (NLIE) and the analytic properties, which enable us to obtain the partition function (and free energy). In Section 5, we have succeeded in giving the associated $t-W$ relations among the transfer matrices, which allow one to derive the associated NLIEs. Taking the $SU(3)$-invariant spin chain an example, we apply our method to  obtain the corresponding free energy. In Section 6, we summarize our results and give some discussions. Some supporting materials are given in Appendices A-F.


\section{Heisenberg chain and the associated  QTM}
\label{XXX} \setcounter{equation}{0}

The integrability of the model (\ref{xxzh})-(\ref{BC}) is associated with the well-known rational six-vertex  $R$-matrix
\bea
R(u)=\lt(\begin{array}{llll}u+\eta&&&\\&u&\eta&\\
&\eta& u&\\&&&u+\eta\end{array}\rt),
\label{r-matrix} \eea
where $u$ is the spectral parameter and the crossing parameter $\eta=i$. The $R$-matrix satisfies the quantum
Yang-Baxter equation (QYBE)
\begin{eqnarray}
\hspace{-1.2truecm}R_{12}(u_1-u_2)R_{13}(u_1-u_3)R_{23}(u_2-u_3)=R_{23}(u_2-u_3)R_{13}(u_1-u_3)R_{12}(u_1-u_2),\label{QYB}
\end{eqnarray}
and the properties:
\bea
&&\hspace{-1.5cm}\mbox{ Initial
condition}:\,R_{12}(0)=  \eta P_{12},\label{Int-R}\\
&&\hspace{-1.5cm}\mbox{ Unitarity
relation}:\,R_{12}(u)R_{21}(-u)= -\xi(u)\times\,{\rm id},
\quad \xi(u)=(u-\eta)(u+\eta),\label{Unitarity}\\
&&\hspace{-1.5cm}\mbox{ Crossing
relation}:\,R_{12}(u)=V_1R_{12}^{t_2}(-u-\eta)V_1,\quad
V=-i\s^y,
\label{crosing-unitarity}\\
&&\hspace{-1.5cm}\mbox{ Fusion condition}:\,R_{12}(\pm\eta)=\pm\eta P^{(\pm)}_{12}=\pm\eta \frac{1\pm P_{12}}{2},
\label{Fusion-con}\\
&&\hspace{-1.5cm}\mbox{ PT-symmetry}:\,R_{12}(u)=R_{21}(u)=R^{t_1\,t_2}_{12}(u),\label{PT}\\
&&\hspace{-1.4cm}\mbox{$Z_2$-symmetry}: \,
\s^i_1\s^i_2R_{12}(u)=R_{12}(u)\s^i_1\s^i_2,\quad
\mbox{for}\,\,
i=x,y,z.\label{Z2-sym}
\eea
Here $R_{21}(u)=P_{12}R_{12}(u)P_{12}$ with $P_{12}$ being
the usual permutation operator and $t_i$ denotes transposition in
the $i$-th space. Throughout this paper we adopt the standard notations: for
any matrix $A\in {\rm End}(\Cb^2)$, $A_j$ is an embedding operator
in the tensor space $\Cb^2\otimes \Cb^2\otimes\cdots$, which acts
as $A$ on the $j$-th space and as identity on the other factor
spaces; $R_{ij}(u)$ is an embedding operator of R-matrix in the
tensor space, which acts as identity on the factor spaces except
for the $i$-th and $j$-th ones.

Let us introduce the transfer matrix $t(u)$ of the XXX closed chain \cite{kor97}
 \begin{eqnarray}
 t^{(L)}(u)=tr_0\{T^{(L)}_0(u)\}=tr_0\lt\{\,R_{0L}(u)\ldots R_{01}(u)\rt\},\label{chain-trans-per}
 \end{eqnarray}
where $tr_0$ denotes trace over the
``auxiliary space" $0$. The expression (\ref{r-matrix}) of the $R$-matrix $R(u)$, the definition (\ref{chain-trans-per}) of the transfer matrix  imply that
\bea
t^{(L)}(u)=2u^L+t^{(1)}u^{L-1}+\cdots+t^{(L-1)}u+t^{(L)}.\no 
\eea
Moreover, the Hamiltonian described by (\ref{xxzh}) and (\ref{BC}) can be expressed in terms of the transfer matrix $t^{(L)}(u)$ as
\begin{eqnarray}
H=H_0-JL+\frac{h}{2}\sum_{j=1}^L\sigma_j^z,\quad H_0=2\eta J \,\frac{\partial \ln t^{(L)}(u)}{\partial
u}|_{u=0},\label{Old-ham}
\end{eqnarray}
which implies that for a small $u$ the transfer matrix has the expansion
\bea
t^{(L)}(u)&=&t^{(L)}(0)\lt(1+\frac{u}{2\eta J}
H_0+O(u^2)\rt)=t^{(L)}(0)e^{\frac{u}{2\eta J}H_0+O(u^2)},\no
\eea
$\hspace{-0.04truecm}$where $t^{(L)}(0)={\eta}^L P_{1L}\cdots P_{12}$. The above relation and the crossing-symmetry (\ref{crosing-unitarity}) of the $R$-matrix allow one to introduce
a quantum transfer matrix $t^{(Q)}(u)$ \cite{klumper1992free,klumper1998free},
\bea
t^{(Q)}(u)&=&tr_0 \{e^{\frac{h\b}{2}\sigma_0^z}\lt(R_{0\,N}(u-\frac{2\eta J\b}{N})\,R_{0\,N-1}(u+\frac{2\eta J\b}{N}-\eta)\rt)\ldots \no\\[4pt] &&\quad\quad\times R_{0\,2}(u-\frac{2\eta J\b}{N})\,R_{0\,1}(u+\frac{2\eta J\b}{N}-\eta)\},\label{Q-trans-per}
\eea
where the positive real parameter $\b$ is related to the temperature $T$ of the system as $\b=\frac{1}{T}$. For a very large even integer $N$, the  partition function $Z(\beta)$ of the spin-$\frac{1}{2}$ XXX closed chain described by the Hamiltonian (\ref{xxzh}) and (\ref{BC}) at a temperature $T$ can be expressed in terms of the quantum transfer matrix $t^{(Q)}(u)$ by the QTM method (for details the reader is referred to Ref.\cite{Ess05}),
\bea
 Z(\b)&=&\lim_{L\rightarrow\infty} tr_{1,\cdots,L}\lt\{e^{-\b H} \rt\}\no\\[4pt]
 &=&e^{\b JL} \lim_{L\rightarrow\infty} tr_{1,\cdots,L}\lt\{\lim_{N\rightarrow\infty}
 \lt\{(1-\frac{2\b}{N} (H+JL)+O(\frac{1}{N^2})\rt\}^{\frac{N}{2}}\rt\}\no\\[4pt]
 &=&e^{\b JL}\lim_{L\rightarrow\infty}\lim_{N\rightarrow\infty}tr_{1,\cdots,N}\lt\{\,\lt\{t^{(Q)}(0)\rt\}^L\rt\}\no\\
 &=&e^{\b JL}\lim_{L\rightarrow\infty}\lim_{N\rightarrow\infty}\lt\{\Lambda^{(Q)}(0)_{max}\rt\}^L.\label{partition-function}
\eea
Here $\Lambda^{(Q)}(0)_{max}$ is the eigenvalue corresponding to the state with the maximus value  $|\Lambda^{(Q)}(0)|$. Moreover, it was shown \cite{Ess05,klumper1992free,klumper1998free} that in the limit of $N\rightarrow \infty$
$\Lambda^{(Q)}(0)_{max}$ is gaped from the others eigenvalues of $\Lambda^{(Q)}(0)$.

\section{T-W relation and eigenvalues of the transfer matrix}
\label{TW} \setcounter{equation}{0}

Similarly as the quantum transfer matrix (\ref{Q-trans-per}), for a large even positive integer $N$, let us introduce another transfer matrix $t(u)$
\bea
&&t(u)=tr_0\lt\{e^{\frac{h\b}{2}\sigma_0^z}\lt(R_{0N}(u-\theta_N)\,R_{0N-1}(u-\theta_{N-1})\cdots R_{02}(u-\theta_2)\,R_{0\,1}(u-\theta_1)\rt)\rt\}, \label{trans-per}
\eea
where $\{\theta_j|j=1,\cdots,N\}$ are some generic complex number, which are called the inhomogeneous parameters (for the special choice of the inhomogeneous parameters, one can recover the quantum transfer matrix (\ref{Q-trans-per})).
The expression (\ref{r-matrix}) of the $R$-matrix $R(u)$, the definition (\ref{trans-per})
of the transfer matrix $t(u)$ imply that
\bea
t(u)=2\cosh\frac{h\b}{2}u^N+t^{(1)}u^{N-1}+\cdots+t^{(N-1)}u+t^{(N)}. \label{Expansion}
\eea
Moreover with the help of the fusion of $R$-matrix \cite{Kir86},
we can derive that the transfer matrix $t(u)$ satisfies  the $t-W$ relation
\bea
t(u)\,t(u-\eta)=a(u)\,d(u-\eta)\times {\rm id}+e^{\frac{h\b}{2}}d(u)\,\mathbb{W}(u),\label{t-W-relation-op}
\eea
where the functions $a(u)$ and $d(u)$ are given by
\bea
a(u)=e^{\frac{h\b}{2}}\prod_{k=1}^N(u-\theta_k+\eta),\quad d(u)=e^{-\frac{h\b}{2}}\prod_{k=1}^N(u-\theta_k).\label{a-d-functions}
\eea
and $\mathbb{W}(u)$ (given by  below (\ref{W-op})), as a function of $u$, is an operator-valued polynomial of degree $N$, which actually is some fused transfer matrix of the fundamental one. The details of the proof the $t-W$ relation (\ref{t-W-relation-op}) will be given in Appendix A.

It is easy to shown that the transfer matrices $t(u)$ and $\mathbb{W}(u)$ commute with each other, namely,
\bea
[t(u),\,t(v)]=[\mathbb{W}(u),\,\mathbb{W}(v)]=[t(u),\,\mathbb{W}(v)]=0,\label{Communtivity}
\eea
which implies that they have common eigenstates. Let $|\Psi\rangle$  be a common eigenstate of the transfer matrices with  eigenvalues $\Lambda(u)$ and $W(u)$, namely,
\bea
t(u)\,|\Psi\rangle=\Lambda(u)\,|\Psi\rangle,\quad \mathbb{W}(u)\,|\Psi\rangle=W(u)\,|\Psi\rangle.
\eea
The operator identity (\ref{t-W-relation-op}) of the transfer matrices  then gives rise to the corresponding relation for their  eigenvalues
\bea
\Lambda(u)\,\Lambda(u-\eta)=a(u)\,d(u-\eta)+e^{\frac{h\b}{2}}d(u)\,W(u).\label{Eigen-id}
\eea
The expansion expression (\ref{Expansion}) and (\ref{Eigen-id}) allow us to express any eigenvalue
$\Lambda(u)$  of the transfer matrix (or $W(u)$ of the fused one) in terms of its $N$ zero points $\{z_j|j=1,\cdots,N\}$ (or $\{w_j|j=1,\cdots,N\}$) as follow
\bea
&&\Lambda(u)=2\cosh\frac{h\b}{2}\prod_{j=1}^{N}\,(u-z_j),\label{Zero-points}\\[4pt]
&&W(u)=(4\cosh^2\frac{h\b}{2}-1)\,\prod_{j=1}^{N}\,(u-w_j).\label{Zero-points-w}
\eea
Taking $u$ at the $2N$ points $\{z_j|j=1,\cdots,N\}$ and $\{w_j|j=1,\cdots,N\}$, we have the associated BAEs
\bea
&&a(z_j)\,d(z_j-\eta)=-e^{\frac{h\b}{2}}d(z_j)\,W(z_j),
\quad j=1,\cdots,N,\label{BAEs-1}\\[4pt]
&&a(w_j)\,d(w_j-\eta)=\Lambda(w_j)\Lambda(w_j-\eta), \quad j=1,\cdots,N.\label{BAEs-2}
\eea
Then $2N$ parameters $\{z_j|j=1,\cdots, N\}$ and $\{w_j|j=1,\cdots,N\}$, which are related to the roots of the eigenvalues $\Lambda(u)$ and $W(u)$, can be determined completely by the above BAEs.

In order to investigate the thermodynamics of the spin-$\frac{1}{2}$ XXX closed chain described by the Hamiltonian (\ref{xxzh}) and (\ref{BC}), let us focus on the quantum transfer matrix  $t^{(Q)}(u)$ given by (\ref{Q-trans-per}) for a large even $N$ and denote its eigenvalue by $\Lambda^{(Q)}(u)$. In this case the inhomogeneous parameters are specially chosen by
(\ref{Q-trans-per}) and the associated functions $a(u)$ and $d(u)$ become
\bea
a(u)\hspace{-0.18truecm}&=&\hspace{-0.18truecm}e^{\frac{h\b}{2}}\lt\{u-\frac{2\eta J\b}{N}+\eta\rt\}^{\frac{N}{2}}\,
\lt\{u+\frac{2\eta J\b}{N}\rt\}^{\frac{N}{2}},\\[4pt]
d(u)\hspace{-0.18truecm}&=&\hspace{-0.18truecm}e^{-\frac{h\b}{2}} \lt\{u-\frac{2\eta J\b}{N}\rt\}^{\frac{N}{2}}\,
\lt\{u+\frac{2\eta J\b}{N}-\eta\rt\}^{\frac{N}{2}}.\label{Q-ad-function}
\eea
The free energy per site $f(\b)$ is given in terms of the partition function (\ref{partition-function}) by
\bea
f(\b)\hspace{-0.18truecm}&=&\hspace{-0.18truecm}-\frac{1}{\b}\lim_{L\rightarrow\infty}\lim_{N\rightarrow\infty}\frac{1}{L}\lt(\ln Z(\b)\rt)\no\\[4pt]
\hspace{-0.18truecm}&=&\hspace{-0.18truecm}-J-\frac{1}{\b}\lim_{L\rightarrow\infty}\lim_{N\rightarrow\infty}\lt\{\ln\Lambda^{(Q)}(0)_{max}\rt\}.\label{Free-energy}
\eea
Hence it is sufficient to calculate $\Lambda^{(Q)}(u)$ of the eigenstate with $|\Lambda^{(Q)}(0)|_{max}$.  Eigenvalues of the QTM  can be also obtained by the
algebraic Bethe ansatz method \cite{kor97} alternatively, where $\Lambda^{(Q)}(u)$ is given in terms of a homogeneous $T-Q$ relation, namely,
\bea
\Lambda^{(Q)}(u)\hspace{-0.2truecm}&=&\hspace{-0.2truecm}a(u)\frac{Q(u-\eta)}{Q(u)}+d(u)\frac{Q(u+\eta)}{Q(u)},\label{T-Q}\\[4pt]
Q(u)\hspace{-0.2truecm}&=&\hspace{-0.2truecm}\prod_{j=1}^M(u-\l_j),\quad M=0,\cdots, N,\no
\eea
where the functions $a(u)$ and $d(u)$ are given in (\ref{Q-ad-function}).
The parameters $\{\l_j|j=1,\cdots,M; M=0,\cdots,N\}$ satisfy the BAEs
\bea
\frac{d(\l_j)}{a(\lambda_j)}=-\frac{Q(\l_j-\eta)}{Q(\l_j+\eta)} ,\quad j=1,\cdots,M.
\eea
It was shown \cite{Ess05,klumper1992free,klumper1998free} that the  eigenvalue of the eigenstate  with $|\Lambda^{(Q)}(0)|_{max}$ belongs to the sector of $M=\frac{N}{2}$ with all the Bethe roots being real.
For the simplicity, let us introduce $M=\frac{N}{2}$ in the following part of the paper, and introduce a parameter $\tau$ (a positive real number ) associated with the temperature as
\bea
\tau=\frac{J\b}{M}=\frac{J}{MT}=\frac{2J\b}{N},\label{tau}
\eea
and a normalized eigenvalue $\bar{\Lambda}^{(Q)}(u)$
\bea
\bar{\Lambda}^{(Q)}(u)=\frac{\Lambda^{(Q)}(u)}{(u-\eta\tau+\eta)^M(u+\eta\tau-\eta)^M}.\label{Normalize}
\eea
The $T-Q$ relation (\ref{T-Q}) allows us to express $\bar{\Lambda}^{(Q)}(u)$  as
\bea
\bar{\Lambda}^{(Q)}(u)=\frac{(u+\eta\tau)^M}{\prod_{k=1}^M(u-\lambda_k)}
\frac{\prod_{k=1}^M(u-\lambda_k-\eta)}{(u+\eta\tau-\eta)^M}+
\frac{(u-\eta\tau)^M}{\prod_{k=1}^M(u-\lambda_k)}
\frac{\prod_{k=1}^M(u-\lambda_k+\eta)}{(u-\eta\tau+\eta)^M},\label{T-Q-1}
\eea
where the real Bethe roots satisfy the associated BAEs
\bea
\frac{(\lambda_j-\eta\tau)^M(\lambda_j+\eta\tau-\eta)^M}{(\lambda_j+\eta\tau)^M(\lambda_j-\eta\tau+\eta)^M}
=-\prod_{k=1}^M\frac{\lambda_j-\lambda_k-\eta}{\lambda_j-\lambda_k+\eta},\quad j=1,\cdots,M.\label{BAEs-3}
\eea

\begin{figure}[htbp]
  \centering
  \subfigure{\label{fig1:subfig:a} 
    \includegraphics[scale=0.6]{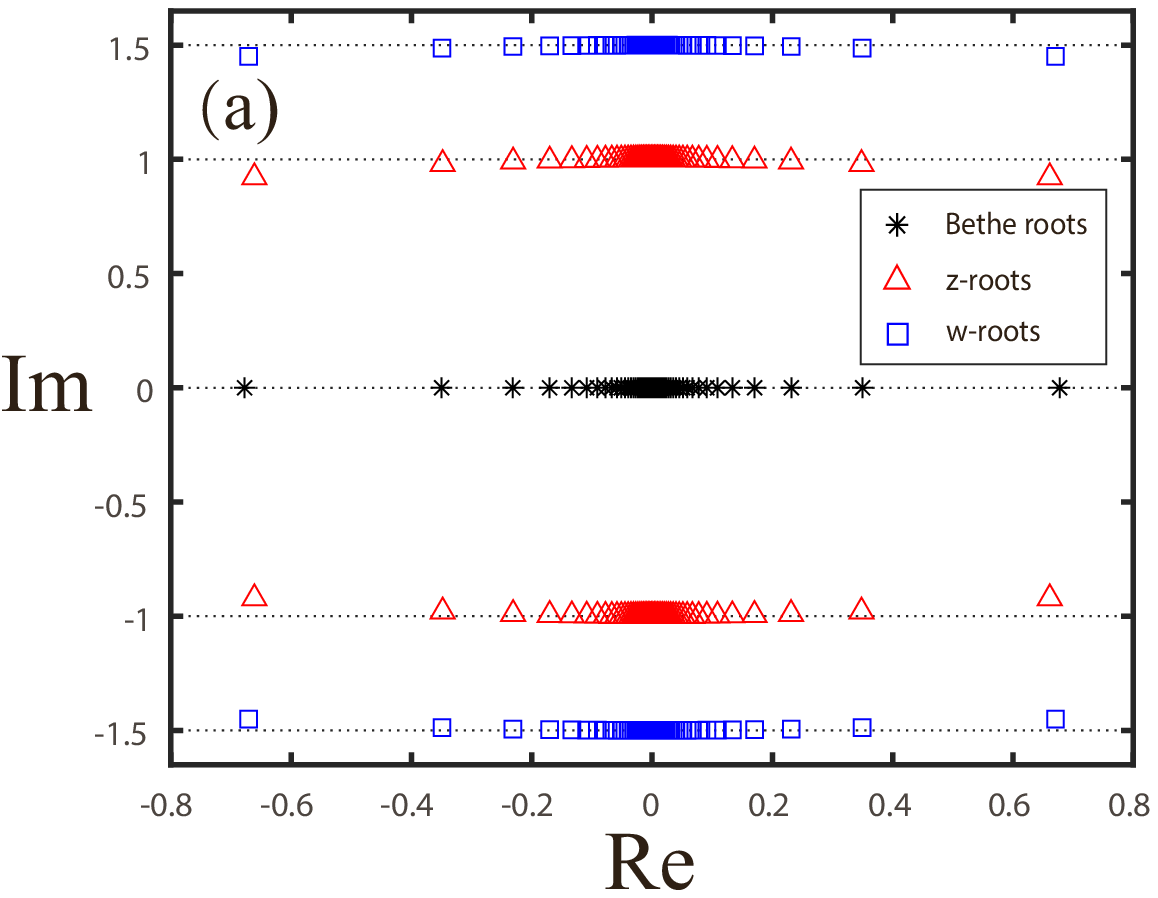}}
  \subfigure{\label{fig1:subfig:b} 
    \includegraphics[scale=0.6]{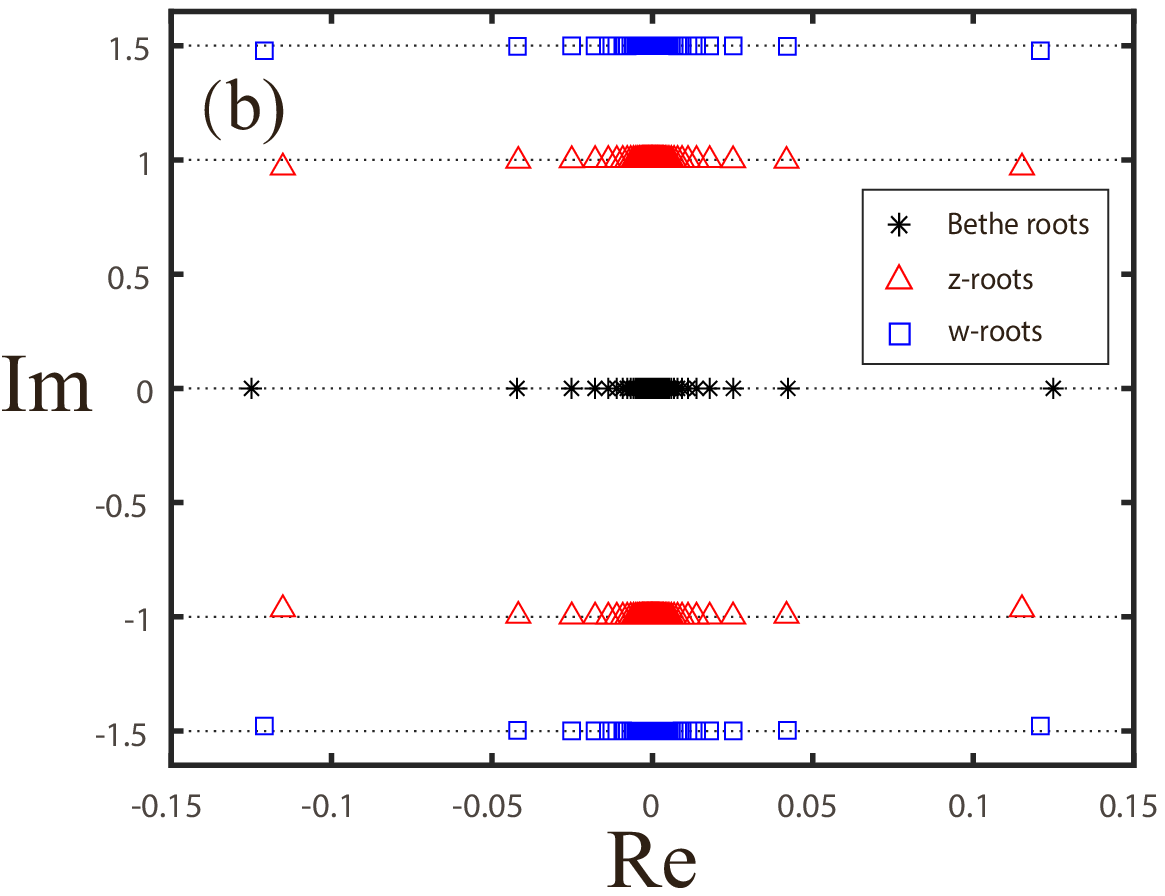}
    }
  \caption{(Colour online) The distributions of the Bethe roots (black $\ast$), $z$-roots (red $\vartriangle$) and $w$-roots (blue $\square$, the roots are shifted down $\frac{\eta}{2}$ corresponding to the zeros of $W^{(Q)}(u+\frac{\eta}{2})$) for the state with the maximus $\Lambda^{(Q)}(0)$ at finite temperatures $T=0.5$ (a) and $T=5$ (b). In both plots, we set $N=100$.}\label{fig1}
\end{figure}
\noindent The distribution of the roots of $\Lambda^{(Q)}(u)$  in figure \ref{fig1} for the state with  $|\Lambda^{(Q)}(0)|_{max}$
for some small $N$ (up to 100) indicates that
the corresponding $\bar{\Lambda}^{(Q)}(u)$ has the decomposition
\bea
\bar{\Lambda}^{(Q)}(u)=2\cosh\frac{h\b}{2}\,\frac{\prod_{j=1}^M(u-u^{(+)}_j-\eta)\,(u-u^{(-)}_j+\eta)}{(u+\eta\tau-\eta)^M\,(u-\eta\tau+\eta)^M},
\label{Zero-points-Z-2}
\eea
where the imaginary parts of $u^{(\pm)}_j$ are close to zero for a large $N$ (namely, ${\rm Im}(u^{(\pm)}_j)\sim 0$) and $\tau$ is given in (\ref{tau}). The Bethe ansatz solution (\ref{T-Q}) also shows that the roots of $\Lambda^{(Q)}(u)$ of the state with  $|\Lambda^{(Q)}(0)|_{max}$ indeed has  the distribution (\ref{Zero-points-Z-2}).
With the help of the $t-W$ relation (\ref{Eigen-id}), we can derive that the eigenvalue $W^{(Q)}(u)$ of the state with   $|\Lambda^{(Q)}(0)|_{max}$ has
the decomposition
\bea
W^{(Q)}(u)=(4\cosh^2\frac{h\b}{2}-1)\,\prod_{j=1}^M(u-w^{(+)}_j-2\eta)(u-w^{(-)}_j+\eta), \label{Zero-points-W-2}
\eea
where ${\rm Im}(w^{(\pm)}_j)\sim 0$ for a large $N$. Then the $t-W$ relation (\ref{Eigen-id}) for the state with the  $|\Lambda^{(Q)}(0)|_{max}$ becomes
\bea
\hspace{-0.68truecm}&&\hspace{-0.68truecm}\quad\quad\bar{\Lambda}^{(Q)}(u\hspace{-0.02truecm}+\hspace{-0.02truecm}\frac{\eta}{2})
 \bar{\Lambda}^{(Q)}(u\hspace{-0.02truecm}-\hspace{-0.02truecm}\frac{\eta}{2})\no\\[8pt]
&&\qquad\qquad= 4\cosh^2\frac{h\b}{2}\frac{\prod_{j=1}^M\hspace{-0.02truecm}(u\hspace{-0.02truecm}-\hspace{-0.02truecm}u^{(-)}_j\hspace{-0.02truecm}+\hspace{-0.02truecm}\frac{3}{2}\eta)
 (u\hspace{-0.02truecm}-\hspace{-0.02truecm}u^{(+)}_j\hspace{-0.02truecm}-\hspace{-0.02truecm}\frac{\eta}{2})
(u\hspace{-0.02truecm}-\hspace{-0.02truecm}u^{(-)}_j\hspace{-0.02truecm}+\hspace{-0.02truecm}\frac{\eta}{2})
(u\hspace{-0.02truecm}-\hspace{-0.02truecm}u^{(+)}_j\hspace{-0.02truecm}-\hspace{-0.02truecm}\frac{3}{2}\eta)}
{(u-\eta\tau+\frac{3}{2}\eta)^M (u+\eta\tau-\frac{\eta}{2})^M (u-\eta\tau+\frac{\eta}{2})^M (u+\eta\tau-\frac{3}{2}\eta)^M}\no\\[8pt]
&&\qquad\qquad=\frac{(u\hspace{-0.02truecm}+\hspace{-0.02truecm}\eta\tau\hspace{-0.02truecm}+\hspace{-0.02truecm}\frac{\eta}{2})^M
(u\hspace{-0.02truecm}-\hspace{-0.02truecm}\eta\tau\hspace{-0.02truecm}-\hspace{-0.02truecm}\frac{\eta}{2})^M}
{(u\hspace{-0.02truecm}-\hspace{-0.02truecm}\eta\tau\hspace{-0.02truecm}+\hspace{-0.02truecm}\frac{\eta}{2})^M
(u\hspace{-0.02truecm}+\hspace{-0.02truecm}\eta\tau\hspace{-0.02truecm}-\hspace{-0.02truecm}
\frac{\eta}{2})^M}\no\\[8pt]
&&\qquad\qquad\qquad\qquad +
(4\cosh^2\frac{h\b}{2}\hspace{-0.02truecm}-\hspace{-0.02truecm}1)
\frac{\prod_{j=1}^M(u\hspace{-0.02truecm}-\hspace{-0.02truecm}w^{(+)}_j\hspace{-0.02truecm}-\hspace{-0.02truecm}\frac{3}{2}\eta)
(u\hspace{-0.02truecm}-\hspace{-0.02truecm}w^{(-)}_j\hspace{-0.02truecm}+\hspace{-0.02truecm}\frac{3}{2}\eta)}
{(u\hspace{-0.02truecm}+\hspace{-0.02truecm}\eta\tau\hspace{-0.02truecm}-\hspace{-0.02truecm}\frac{3}{2}\eta)^M
(u\hspace{-0.02truecm}-\hspace{-0.02truecm}\eta\tau\hspace{-0.02truecm}+\hspace{-0.02truecm}\frac{3}{2}\eta)^M}\no\\[8pt]
&&\qquad\qquad\stackrel{{\rm def}}{=}q(u)
+(4\cosh^2\frac{h\b}{2}\hspace{-0.02truecm}-\hspace{-0.02truecm}1)\bar{w}(u)+O(\frac{1}{N}),\label{t-W-relation-Eig}
\eea
where the functions $q(u)$ and $\bar{w}(u)$ are
\bea
q(u)\hspace{-0.18truecm}&=&\hspace{-0.18truecm}\lim_{M\rightarrow \infty}\,
\frac{(u\hspace{-0.02truecm}+\hspace{-0.02truecm}\eta\tau\hspace{-0.02truecm}+\hspace{-0.02truecm}\frac{\eta}{2})^M
(u\hspace{-0.02truecm}-\hspace{-0.02truecm}\eta\tau\hspace{-0.02truecm}-\hspace{-0.02truecm}\frac{\eta}{2})^M}
{(u\hspace{-0.02truecm}-\hspace{-0.02truecm}\eta\tau\hspace{-0.02truecm}+\hspace{-0.02truecm}\frac{\eta}{2})^M
(u\hspace{-0.02truecm}+\hspace{-0.02truecm}\eta\tau\hspace{-0.02truecm}-\hspace{-0.02truecm}\frac{\eta}{2})^M}=
e^{\frac{2J\b}{u^2+\frac{1}{4}}},\label{q-function}\\[8pt]
\bar{w}(u)\hspace{-0.18truecm}&=&\hspace{-0.18truecm}\lim_{M\rightarrow \infty}\,
\frac{\prod_{j=1}^M(u\hspace{-0.02truecm}-\hspace{-0.02truecm}w^{(+)}_j\hspace{-0.02truecm}-\hspace{-0.02truecm}\frac{3}{2}\eta)
(u\hspace{-0.02truecm}-\hspace{-0.02truecm}w^{(-)}_j\hspace{-0.02truecm}+\hspace{-0.02truecm}\frac{3}{2}\eta)}
{(u\hspace{-0.02truecm}+\hspace{-0.02truecm}\eta\tau\hspace{-0.02truecm}-\hspace{-0.02truecm}\frac{3}{2}\eta)^M
(u\hspace{-0.02truecm}-\hspace{-0.02truecm}\eta\tau\hspace{-0.02truecm}+\hspace{-0.02truecm}\frac{3}{2}\eta)^M}
\stackrel{{\rm def}}{=}e^{-\b\, \bar{\e}(u)}.\label{w-function}
\eea
It is remarked that the function $\bar{\e}(u)$ satisfies the analytic property:
\bea
\hspace{-0.26truecm}\bar{\e}(u) {\rm ~is~ analytic~ except~ some~ singularities~ on~ the~ axis}~ {\rm Im}(u)=\pm\frac{3}{2}~{\rm and}
\, \lim_{u\rightarrow \infty}\bar{\e}(u)=0.\label{Property-epsion-function-1}
\eea

\section{Nonlinear integral equations and the free energy}
\label{FREE} \setcounter{equation}{0}
The decomposition (\ref{Zero-points-Z-2}) and the  very $t-W$ relation (\ref{t-W-relation-Eig}) allow us to give an integral representation
of $\bar{\Lambda}^{(Q)}(u)$ of the state with $|\Lambda^{(Q)}(0)|_{max}$
\bea
\ln \bar{\Lambda}^{(Q)}(u)\hspace{-0.18truecm}&=&\hspace{-0.18truecm}\ln 2\cosh\frac{h\b}{2}+\frac{1}{2\pi i}\oint _{\mathcal{C}_1}dv \,\frac{\ln\lt((q(v)+(4\cosh^2\frac{h\b}{2}-1)
e^{-\b \bar{\e}(v)})/4\cosh^2\frac{h\b }{2}\rt)}{u-v-\frac{\eta}{2}}\no\\[8pt]
&&\qquad\quad\quad\quad +\frac{1}{2\pi i}\oint _{\mathcal{C}_2}dv \,\frac{\ln\lt((q(v)+(4\cosh^2\frac{h\b}{2}-1)
e^{-\b \bar{\e}(v)})/4\cosh^2\frac{h\b }{2}\rt)}{u-v+\frac{\eta}{2}},
\label{Integral-rep-1}
\eea
where the closed integral contour $\mathcal{C}_1$ is surrounding the axis of ${\rm Im}(v)=\frac{1}{2}$, while $\mathcal{C}_2$ is surrounding the axis of ${\rm Im}(v)=-\frac{1}{2}$.   With the help of the $t-W$ relation (\ref{t-W-relation-Eig}) and the integral representation (\ref{Integral-rep-1}), we can derive a NLIE of the function $\bar{\e}(u)$
\bea
\hspace{-0.8truecm}&&\hspace{-0.8truecm}\ln(q(u) \hspace{-0.02truecm}+\hspace{-0.02truecm}(4\cosh^2\frac{h\b}{2}-1)e^{-\b \bar{\e}(u)})=2\ln2\cosh\frac{h\b}{2}\no\\[8pt]
\hspace{-1.2truecm}&&+\frac{1}{2\pi i}\hspace{-0.02truecm}\oint _{\mathcal{C}_1}\hspace{-0.02truecm}dv \hspace{-0.02truecm}(\frac{1}{u\hspace{-0.02truecm}-\hspace{-0.02truecm}v}
\hspace{-0.02truecm}+\hspace{-0.02truecm}\frac{1}{u\hspace{-0.02truecm}-\hspace{-0.02truecm}v\hspace{-0.02truecm}-\hspace{-0.02truecm}\eta})
\hspace{-0.02truecm}\ln\lt((q(v)\hspace{-0.02truecm}+\hspace{-0.02truecm}(4\cosh^2\frac{h\b}{2}\hspace{-0.02truecm}-\hspace{-0.02truecm}1)
e^{-\b \bar{\e}(v)})/4\cosh^2\frac{h\b }{2}\rt)\no\\[8pt]
\hspace{-1.2truecm}&&+\frac{1}{2\pi i}\hspace{-0.02truecm}\oint _{\mathcal{C}_2}\hspace{-0.02truecm}dv \hspace{-0.02truecm}(\frac{1}{u\hspace{-0.02truecm}-\hspace{-0.02truecm}v\hspace{-0.02truecm}+\hspace{-0.02truecm}\eta}
\hspace{-0.02truecm}+\hspace{-0.02truecm}\frac{1}{u\hspace{-0.02truecm}-\hspace{-0.02truecm}v})
\hspace{-0.02truecm}\ln\lt((q(v)\hspace{-0.02truecm}+\hspace{-0.02truecm}(4\cosh^2\frac{h\b}{2}\hspace{-0.02truecm}-\hspace{-0.02truecm}1)
e^{-\b \bar{\e}(v)})/4\cosh^2\frac{h\b }{2}\rt).\label{Integral-Main}
\eea

Due to the fact that the roots and the poles of $ \bar{\Lambda}^{(Q)}(u)$ locate nearly on the two lines with imaginary parts close to $\pm 1$ (see the decomposition (\ref{Zero-points-Z-2})),
we can use the Fourier transformation to obtain another integral representation of $\bar{\Lambda}^{(Q)}(u)$
\bea
\ln \bar{\Lambda}^{(Q)}(u)\hspace{-0.18truecm}&=&\hspace{-0.18truecm}\int_{-\infty}^{+\infty}\frac{dv}{2\cosh\pi(u-v)}\lt\{\ln\bar{\Lambda}^{(Q)}(v+\frac{\eta}{2})+
\ln\bar{\Lambda}^{(Q)}(v-\frac{\eta}{2}) \rt\}\no\\[8pt]
\hspace{-0.18truecm}&=&\hspace{-0.18truecm}\int_{-\infty}^{+\infty}\frac{dv}{2\cosh\pi(u-v)}\lt\{\frac{2J\b}{v^2+\frac{1}{4}}+\ln(1
+(4\cosh^2\frac{h\b}{2}-1)q^{-1}(v)\bar{w}(v))\rt\}.
\eea
where we have used $\eta=i$. Let us introduce the dressing energy function $\e(u)$
\bea
\e(u)=-\frac{1}{\b}\ln \lt(q^{-1}(u)\bar{w}(u)\rt)=\frac{2J}{u^2+\frac{1}{4}}+\bar{\e}(u),\quad \lim_{u\rightarrow\infty}\e(u)=0. \label{epsilon-function}
\eea
It is believed that the analytic property (\ref{Property-epsion-function-1}) and the NLIE (\ref{Integral-Main}) and the asymptotical behavior (\ref{epsilon-function})  might completely determine the function $\bar{\e}(u)$.

Finally we obtain the free energy of the XXX chain described by the
Hamiltonian (\ref{xxzh})-(\ref{BC}) as
\bea
f(\beta)\hspace{-0.18truecm}&=&\hspace{-0.18truecm}J-\frac{1}{\b}\ln \bar{\Lambda}^{(Q)}(0)\no\\[8pt]
\hspace{-0.18truecm}&=&\hspace{-0.18truecm}J-J\int_{-\infty}^{+\infty}
\frac{dv}{\cosh\pi v}\frac{1}{v^2+\frac{1}{4}}
-\frac{1}{\b} \int_{-\infty}^{+\infty}\frac{dv}
{2\cosh\pi v}
\ln \lt(1+(4\cosh^2\frac{h\b}{2}-1)
e^{-\b \e(v)}\rt)\no\\[8pt]
\hspace{-0.18truecm}&=&\hspace{-0.18truecm}e_g-\frac{1}{\b} \int_{-\infty}^{+\infty}\frac{dv}{2\cosh\pi v}\,
\ln \lt(1+(4\cosh^2\frac{h\b}{2}-1)e^{-\b \e(v)}\rt).\label{free energy}
\eea
where $e_g=\hspace{-0.18truecm}J-J\int_{-\infty}^{+\infty}
\frac{dv}{\cosh\pi v}\frac{1}{v^2+\frac{1}{4}}$ is the energy of the ground state for the XXX chain (\ref{xxzh})-(\ref{BC}) \cite{takahashi2005thermodynamics} and the dressing energy
$\e(u)$ satisfying the relations (\ref{Property-epsion-function-1}), (\ref{Integral-Main}) and (\ref{epsilon-function}).

\begin{figure}[htbp]
\centering
\includegraphics[scale=0.6]{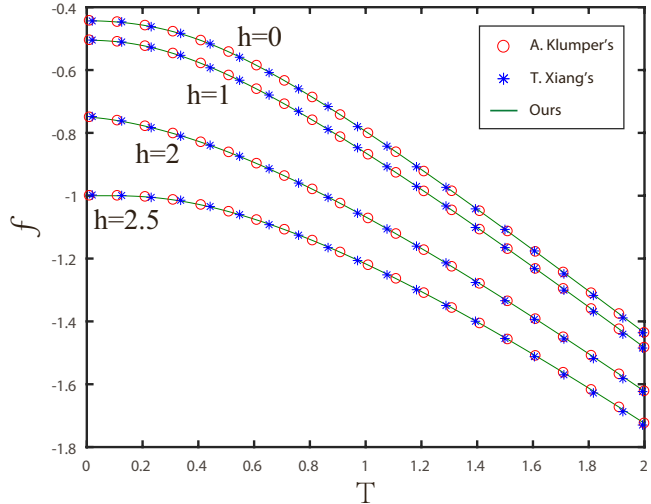}
  \caption{Free energy $f$ vs $T$ for the closed XXX chain in different magnetic fields.}\label{fig2}
\end{figure}

Using the numerical iterative procedure in Appendix B, we obtain the free energy $f$ variation with temperature $T$ in different magnetic fields as shown in figure \ref{fig2}. From the figure,
we find  that our result coincides well with the those of \cite{klumper1992free,klumper1998free} and \cite{PhysRevB.58.9142} obtained with  different approaches.
Moreover, the analytic property (\ref{Property-epsion-function-1}) and the NLIE (\ref{Integral-Main}) allow us to give
the HTE of the free energy as
\bea
f/T=-\ln(2\cosh(h/T))-\frac{J}{T}\tanh^2(h/T)-\frac{3J^2}{2T^2}(1-\tanh^4(h/T))+\cdot\cdot\cdot,\label{HTE-XXX}
\eea
which recovers that of \cite{PhysRevLett.89.117201} obtained previously with a different approach. The details of the derivation of (\ref{HTE-XXX}) will be given in Appendix C.

Our method to obtain the NLIEs is more direct and easily extensible to other quantum integrable  spin chains associated other Lie algebras. We will apply the method to construct the corresponding NLIEs for the $SU(n)$ quantum spin chain in the next section. Moreover, the procedure to obtain the NLIEs  does not directly  depend on the Bethe ansatz solution (\ref{T-Q}) and (\ref{BAEs-3}) of the model, which  is related only to  the patterns of the root distributions  of $\Lambda^{Q}(u)$ and the fused ones. The roots distributions can be obtained by directly  solving  the equations (\ref{BAEs-1})-(\ref{BAEs-2}) (or the below equations (\ref{BAEs-su3-1})-(\ref{BAEs-su3-4}) for the $SU(3)$-case).
\section{$SU(n)$-invariant spin chain}
\label{su3case}\setcounter{equation}{0}

Let \textbf{V} denote an $n$-dimensional linear space. The Hamiltonian of $SU(n)$-invariant quantum spin system on a $L$-sites lattice with the periodic boundary condition is given by \cite{kor97}
\begin{eqnarray}
\qquad H = J\sum_{j=1}^LP_{jj+1},\qquad P_{jj+1}=\sum_{\mu,\nu=1}^nE_j^{\mu,\nu}E_{j+1}^{\nu,\mu},\label{su-n}
\end{eqnarray}
where $P_{jj+1}$ is permutation operator and $(E^{\mu,\nu})_{\beta}^{\alpha}=\delta_{\alpha}^{\mu}\delta_{\nu}^{\beta}$ with $\alpha,\beta,\gamma,\delta=1,\cdot\cdot\cdot,n$. The integrability of the system (\ref{su-n}) is guaranteed by the $SU(n)$-invariant $R$-matrix $R(u)\in {\rm End}(\textbf{V}\otimes\textbf{V})$ \cite{BS1975,Perk81}.
\begin{eqnarray}
R_{12}(u)=u+\eta P_{12}.  \label{r-matrix-su-n}
\end{eqnarray}
Besides the QYBE, the $R$-matrix satisfies the properties:
\bea
&&\hspace{-1.5cm}\mbox{ Initial
condition}:\,R_{12}(0)=\eta P_{12},\label{Int-R-su-n}\\
&&\hspace{-1.5cm}\mbox{ Unitarity
relation}:\,R_{12}(u)R_{21}(-u)= -\rho_1(u)\times\,{\rm id},
\quad \rho_1(u)=(u-\eta)(u+\eta),\label{Unitarity-su-n}\\
&&\hspace{-1.5cm}\mbox{ Crossing-unitarity}:\,R_{12}^{t_1}(u)R_{21}^{t_1}(-u-n\eta)=-\rho_2(u)\times\,{\rm id},
\quad \rho_2(u)=u(u+3\eta),
\label{crosing-unitarity-su-n}\\
&&\hspace{-1.5cm}\mbox{ PT-symmetry}:\,R_{12}(u)=R_{21}(u)=R^{t_1\,t_2}_{12}(u),\label{PT-su-n}\\
&&\hspace{-1.5cm}\mbox{ Fusion conditions}:\,R_{12}(-\eta)=-2\eta P_{12}^{(-)},\quad R_{12}(\eta)=2\eta P_{12}^{(+)}.\label{FC-su-n}
\eea
The corresponding QTM can be constructed as follow \cite{FUJII1999}
\bea
t_1^{(Q)}(u)=tr_0\lt\{e^{\frac{h\beta}{2}S_0}\lt(R_{0N}(u-\eta\tau)R_{N-1\,0}^{t_{N-1}}(u+\eta\tau)\rt)\cdot\cdot\cdot
\lt(R_{02}(u-\eta\tau)R_{10}^{t_1}(u+\eta\tau)\rt)\rt\},\label{t-Q1-su-n}
\eea
where the diagonal matrix $S_0={\rm diag}(\mu_1,\mu_2,...,\mu_n)$ is related to the external field and $\tau$ is given in (\ref{tau}).  In the case of the $SU(3)$ invariant spin chain, we have $S_0=S_0^z={\rm diag}(1,0,-1)$. The expression (\ref{r-matrix-su-n}) of the $R$-matrix $R(u)$, the definition (\ref{t-Q1-su-n}) of the QTM imply that
\bea
t_1^{(Q)}(u)=(\sum_{j=1}^n e^{\frac{h\beta}{2}\mu_j})u^N+t_1^{(1)}u^{N-1}+\cdots+t_1^{(N-1)}u+t_1^{(N)}. \label{Expansion-su-n}
\eea
With the help of the fusion \cite{Kul81,Kir86} of the $R$-matrix we can introduce some fused quantum transfer matrices\footnote{It is remarked that the fused transfer matrices $\{t_i^{(Q)}(u)|i=2,\cdot\cdot\cdot,n-1\}$ in this paper correspond to those
$\{\tau_i^{(p)}(u)|i=2,\cdot\cdot\cdot,n-1\}$ in \cite{Cao14}.}  $\{t_i^{(Q)}(u)|i=2,\cdot\cdot\cdot,n-1\}$ which are related to the representations
associated with the other $n-2$ fundamental highest weights of $su(n)$ algebra  and their counterparts $\{\mathbb{W}_{i}^{(Q)}(u)|i=1,\cdot\cdot\cdot,n-1\}$.
Using the method developed in \cite{Cao14} we can derive that the fused transfer matrices satisfy the associated $t-W$ relations
\bea
t_m^{(Q)}(u)\,t_m^{(Q)}\hspace{-0.7cm}&&(u-\eta)=t_{m-1}^{(Q)}(u-\eta)\,t_{m+1}^{(Q)}(u)+a_m(u)\mathbb{W}_m^{(Q)}(u), \qquad m=1,\cdot\cdot\cdot,n-1.\label{t-W-su(n)}
\eea
The operators $t_0^{(Q)}(u)$, $t_n^{(Q)}(u)$ and the functions $a_m(u)$ are given by
\bea
&&t_0^{(Q)}(u)=[(u-\eta\tau-\eta)(u+\eta\tau)]^{\frac{N}{2}}\times{\rm id},\quad t_n^{(Q)}(u)=[(u+\eta\tau-n\eta)(u-\eta\tau+\eta)]^{\frac{N}{2}}\times{\rm id},\\[6pt]
 &&a_m(u)=[(u+\eta\tau-m\eta)(u-\eta\tau)]^{\frac{N}{2}}, \qquad m=1,\cdot\cdot\cdot,n-1.\label{t-W-su(n)-1}
\eea
The proof of  the relations (\ref{t-W-su(n)}) will be given in Appendix A.

Some remarks are in order. We have introduced the $2n-3$ extra (or auxiliary) fused transfer matrices  $\{t^{(Q)}_i(u)|i=2,\cdot\cdot\cdot,n-1\}$ and $\{\mathbb{W}_{i}^{(Q)}(u)|i=1,\cdot\cdot\cdot,n-1\}$.
Hence in order to determine the eigenvalue $\Lambda^{(Q)}(u)$ of the original quantum transfer  matrix $t^{(Q)}(u)$, we need to further introduce $2n-3$ auxiliary functions (c.f., $2^n-2$ auxiliary functions for the $SU(n)$ case \cite{FUJII1999,Dam06}) which correspond to the  eigenvalues of the resulting fused transfer matrices.

\subsection{T-W relations of the $SU(3)$-variant chain}
\label{su3-1}

Taking the $SU(3)$-invariant spin chain as an example, we shall show how our method works in the following parts of the section.
The Hamiltonian of the $SU(3)$-invariant closed spin chain is given by
\begin{eqnarray}
\qquad H = J\sum_{j=1}^LP_{j,j+1},\qquad P_{j,j+1}=\sum_{\mu,\nu=1}^3E_j^{\mu,\nu}E_{j+1}^{\nu,\mu},  \label{su3}
\end{eqnarray}
with the periodic boundary condition
\bea
E_{L+1}^{\mu,\nu}=E_{1}^{\mu,\nu},\quad {\rm  for}\quad \mu,\nu=1,2,3.\label{BC-su3}
\eea
The associated $R$-matrix reads
\bea
 R(u)=\lt(
            \begin{array}{ccc|ccc|ccc}
              u+\eta &  &  &  &  &  &  &  &  \\
               & u &  & \eta &  &  &  &  &  \\
               &  & u &  &  &  & \eta &  &  \\\hline
               & \eta &  & u &  &  &  &  &  \\
               &  &  &  & u+\eta &  &  &  &  \\
               &  &  &  &  & u &  & \eta &  \\\hline
               &  & \eta &  &  &  & u &  &  \\
               &  &  &  &  & \eta &  & u &  \\
               &  &  &  &  &  &  &  & u+\eta \\
            \end{array}
 \rt),
\label{r-matrix-su3} \eea

For the periodic $SU(3)$ model with an external field $h$, the corresponding QTM can be constructed as follow \cite{FUJII1999}
\bea
t_1^{(Q)}(u)=tr_0\lt\{e^{\frac{h\beta}{2}S^z_0}\lt(R_{0N}(u-\eta\tau)R_{N-1\,0}^{t_{N-1}}(u+\eta\tau)\rt)\cdot\cdot\cdot
\lt(R_{02}(u-\eta\tau)R_{10}^{t_1}(u+\eta\tau)\rt)\rt\},\label{t-Q1-su3}
\eea
where the operator $S^z_0={\rm diag}(1,0,-1)$. The expression (\ref{r-matrix-su3}) of the $R$-matrix $R(u)$, the definition (\ref{t-Q1-su3})  of the QTM imply that
\bea
t_1^{(Q)}(u)=(2\cosh\frac{h\b}{2}+1)u^N+t_1^{(1)}u^{N-1}+\cdots+t_1^{(N-1)}u+t_1^{(N)}. \label{Expansion-su3}
\eea
The corresponding $t-W$ relations\footnote{For later calculative and notational convenience, we shift the spectral parameter $u$ of the transfer matrix $t_2^{(Q)}(u)$ to be $u-\frac{\eta}{2}$ in the first $t-W$ relation (\ref{t-W-relation-su31}).} (\ref{t-W-su(n)}) read
\begin{small}
\bea
\hspace{-0.38cm}&&t_1^{(Q)}(u)t_1^{(Q)}(u\hspace{-0.06cm}-\hspace{-0.06cm}\eta)\hspace{-0.06cm}=\hspace{-0.06cm}[(u\hspace{-0.06cm}+\hspace{-0.06cm}\eta\tau)(u\hspace{-0.06cm}-\hspace{-0.06cm}\eta\tau-\eta)]^{\frac{N}{2}}t_2^{(Q)}
(u\hspace{-0.06cm}-\hspace{-0.06cm}
\frac{\eta}{2})\hspace{-0.06cm}+\hspace{-0.06cm}[(u\hspace{-0.06cm}+\hspace{-0.06cm}\eta\tau\hspace{-0.06cm}-\hspace{-0.06cm}\eta)(u\hspace{-0.06cm}-\hspace{-0.06cm}\eta\tau)]^{\frac{N}{2}}\mathbb{W}^{(Q)}_1(u)\label{t-W-relation-su31},\\[6pt]
\hspace{-0.38cm}&&t_2^{(Q)}(u)t_2^{(Q)}(u\hspace{-0.06cm}-\hspace{-0.06cm}\eta)\hspace{-0.06cm}=\hspace{-0.06cm}[(u\hspace{-0.06cm}+\hspace{-0.06cm}\eta\tau\hspace{-0.06cm}-\hspace{-0.06cm}\frac{5}{2}\eta)(u\hspace{-0.06cm}
-\hspace{-0.06cm}\eta\tau\hspace{-0.06cm}+\hspace{-0.06cm}\frac{3}{2}\eta)]^{\frac{N}{2}}
t_1^{(Q)}(u\hspace{-0.06cm}-\hspace{-0.06cm}\frac{\eta}{2})\hspace{-0.06cm}+
\hspace{-0.06cm}[(u\hspace{-0.06cm}+\hspace{-0.06cm}\eta\tau\hspace{-0.06cm}-\hspace{-0.06cm}\frac{3}{2}\eta)(u\hspace{-0.06cm}-\hspace{-0.06cm}\eta\tau\hspace{-0.06cm}+\hspace{-0.06cm}\frac{\eta}{2})]^{\frac{N}{2}}\mathbb{W}^{(Q)}_2(u).
\label{t-W-relation-su32}
\eea
\end{small}
\hspace{-0.1cm}The resulting transfer matrices $t_2^{(Q)}(u)$, $\mathbb{W}_1^{(Q)}(u)$ and $\mathbb{W}_2^{(Q)}(u)$, as the functions of $u$, are three operator-valued polynomials of degree $N$. In addition, the transfer matrices $t_1^{(Q)}(u)$, $t_2^{(Q)}(u)$, $\mathbb{W}_1^{(Q)}(u)$ and $\mathbb{W}_2^{(Q)}(u)$  commute with each other,
\bea
[t_i^{(Q)}(u),\,t_j^{(Q)}(v)]=[\mathbb{W}_i^{(Q)}(u),\,\mathbb{W}_j^{(Q)}(v)]=[t_i^{(Q)}(u),\,\mathbb{W}_j^{(Q)}(v)]=0,\quad i,j=1,2.\label{Communtivity-su3}
\eea
The commutativity (\ref{Communtivity-su3}) of the transfer matrices $t_1^{(Q)}(u)$, $t_2^{(Q)}(u)$, $\mathbb{W}_1^{(Q)}(u)$ and $\mathbb{W}_2^{(Q)}(u)$ with different spectral parameters implies that they have common eigenstates. Let $|\Psi\rangle$  be a common eigenstate of the QTMs with the eigenvalues $\Lambda_1^{(Q)}(u)$, $\Lambda_2^{(Q)}(u)$, $W_1^{(Q)}(u)$ and $W_2^{(Q)}(u)$, namely
\bea
t_i^{(Q)}(u)\,|\Psi\rangle=\Lambda_i^{(Q)}(u)\,|\Psi\rangle,\quad \mathbb{W}_i^{(Q)}(u)\,|\Psi\rangle=W_i^{(Q)}(u)\,|\Psi\rangle,\quad i=1,2.\no
\eea
The operator identities (\ref{t-W-relation-su31}) and (\ref{t-W-relation-su32}) of the QTMs then give rise to the corresponding relations for their
eigenvalues
\begin{small}
\bea
\hspace{-0.38truecm}&&\Lambda_1^{(Q)}(u)\Lambda_1^{(Q)}(u\hspace{-0.06cm}-\hspace{-0.06cm}\eta)\hspace{-0.06cm}=\hspace{-0.06cm}[(u\hspace{-0.06cm}+\hspace{-0.06cm}\eta\tau)(u\hspace{-0.06cm}-\hspace{-0.06cm}\eta\tau\hspace{-0.06cm}-\hspace{-0.06cm}\eta)]^{\frac{N}{2}}
\Lambda_2^{(Q)}(u\hspace{-0.06cm}-\hspace{-0.06cm}\frac{\eta}{2})\hspace{-0.06cm}+\hspace{-0.06cm}[(u\hspace{-0.06cm}+\hspace{-0.06cm}\eta\tau\hspace{-0.06cm}-\hspace{-0.06cm}\eta)(u\hspace{-0.06cm}-\hspace{-0.06cm}\eta\tau)]^{\frac{N}{2}}W_1^{(Q)}(u)\label{Eigen-id-su31},\\
\hspace{-0.38truecm}&&\Lambda_2^{(Q)}(u)\Lambda_2^{(Q)}(u\hspace{-0.06cm}-\hspace{-0.06cm}\eta)\hspace{-0.06cm}=\hspace{-0.06cm}[(u\hspace{-0.06cm}+\hspace{-0.06cm}\eta\tau\hspace{-0.06cm}-\hspace{-0.06cm}\frac{5}{2}\eta)(u\hspace{-0.06cm}-\hspace{-0.06cm}\eta\tau\hspace{-0.06cm}
+\hspace{-0.06cm}\frac{3}{2}\eta)]^{\frac{N}{2}}\Lambda_1^{(Q)}(u\hspace{-0.06cm}-\hspace{-0.06cm}\frac{\eta}{2})\hspace{-0.06cm}+\hspace{-0.06cm}[(u\hspace{-0.06cm}+\hspace{-0.06cm}\eta\tau\hspace{-0.06cm}-\hspace{-0.06cm}\frac{3}{2}\eta)(u\hspace{-0.06cm}-\hspace{-0.06cm}\eta\tau\hspace{-0.06cm}+\hspace{-0.06cm}\frac{\eta}{2})]^{\frac{N}{2}}W_2^{(Q)}(u)\label{Eigen-id-su32}.
\eea
\end{small}
\hspace{-0.1cm}The expansion expression (\ref{Expansion-su3}), (\ref{Eigen-id-su31}) and (\ref{Eigen-id-su32}) allow us to express any eigenvalue
$\Lambda_1^{(Q)}(u)$  (or $\Lambda_2^{(Q)}(u)$, $W_1^{(Q)}(u)$ and $W_2^{(Q)}(u)$) of the QTM in terms of its $N$ zero points $\{z_j^{(1)}|j=1,\cdots,N\}$ (or $\{z_j^{(2)}|j=1,\cdots,N\}$, $\{w_j^{(1)}|j=1,\cdots,N\}$ and $\{w_j^{(2)}|j=1,\cdots,N\}$) as follow
\bea
&&\Lambda_i^{(Q)}(u)=b(\beta)\prod_{j=1}^{N}\,(u-z^{(i)}_j),\quad W_i^{(Q)}(u)=(b^2(\beta)-b(\beta))\,\prod_{j=1}^{N}\,(u-w^{(i)}_j),\quad i=1,2,\label{Zero-points-su3}
\eea
where $b(\beta)=2\cosh{\frac{h\beta}{2}}+1$. Taking $u$ at the $4N$ points $\{z_j^{(i)}|j=1,\cdots,N\}$ and $\{w_j^{(i)}|j=1,\cdots,N\} (i=1,2)$, we have the associated BAEs
\begin{footnotesize}
\bea
&&\hspace{-0.4cm}[(z_j^{(1)}+\eta\tau)(z_j^{(1)}-\eta\tau-\eta)]^{\frac{N}{2}}\Lambda_2^{(Q)}(z_j^{(1)}-\frac{\eta}{2})=-[(z_j^{(1)}+\eta\tau-\eta)(z_j^{(1)}-\eta\tau)]^{\frac{N}{2}}W_1^{(Q)}(z_j^{(1)}),
\quad \hspace{-0.12cm}j\hspace{-0.06cm}=\hspace{-0.06cm}1,\cdots,N,\label{BAEs-su3-1}\\[4pt]
&&\hspace{-0.4cm}[(w_j^{(1)}+\eta\tau)(w_j^{(1)}-\eta\tau-\eta)]^{\frac{N}{2}}\Lambda_2^{(Q)}(w_j^{(1)}-\frac{\eta}{2})=\Lambda_1^{(Q)}(w_j^{(1)})\Lambda_1^{(Q)}(w_j^{(1)}-\eta), \quad \hspace{-0.12cm}j\hspace{-0.06cm}=\hspace{-0.06cm}1,\cdots,N,\label{BAEs-su3-2}\\[4pt]
&&\hspace{-0.4cm}[(z_j^{(2)}\hspace{-0.06cm}+\hspace{-0.06cm}\eta\tau\hspace{-0.06cm}-\hspace{-0.06cm}\frac{5}{2}\eta)(z_j^{(2)}\hspace{-0.06cm}-\hspace{-0.06cm}\eta\tau\hspace{-0.06cm}+\hspace{-0.06cm}\frac{3}{2}\eta)]^{\frac{N}{2}}\Lambda_1^{(Q)}(z_j^{(2)}\hspace{-0.06cm}-\hspace{-0.06cm}\frac{\eta}{2})\hspace{-0.06cm}
=\hspace{-0.06cm}-\hspace{-0.02cm}[(z^{(2)}_j\hspace{-0.06cm}+\hspace{-0.06cm}\eta\tau\hspace{-0.06cm}-\hspace{-0.06cm}\frac{3}{2}\eta)(z^{(2)}_j\hspace{-0.06cm}-\hspace{-0.06cm}\eta\tau+\frac{\eta}{2})]^{\frac{N}{2}}W_2^{(Q)}(z^{(2)}_j),
\quad \hspace{-0.12cm}j\hspace{-0.06cm}=\hspace{-0.06cm}1,\cdots,N,\label{BAEs-su3-3}\\[4pt]
&&\hspace{-0.4cm}[(w_j^{(2)}+\eta\tau-\frac{5}{2}\eta)(w_j^{(2)}-\eta\tau+\frac{3}{2}\eta)]^{\frac{N}{2}}\Lambda_1^{(Q)}(w_j^{(2)}-\frac{\eta}{2})=\Lambda_2^{(Q)}(w_j^{(2)})\Lambda_2^{(Q)}(w_j^{(2)}-\eta), \quad \hspace{-0.12cm}j\hspace{-0.06cm}=\hspace{-0.06cm}1,\cdots,N.\label{BAEs-su3-4}
\eea
\end{footnotesize}
\hspace{-0.1cm}Then $4N$ parameters $\{z_j^{(i)}|j=1,\cdots,N\}$ and $\{w_j^{(i)}|j=1,\cdots,N\} (i=1,2)$, which are related to the roots of the eigenvalues $\{\Lambda^{(Q)}_i(u), W^{(Q)}_i(u)|i=1,2\}$, can be determined completely by the above BAEs.
\begin{figure}[htbp]
  \centering
  \subfigure{\label{fig3:subfig:a} 
    \includegraphics[scale=0.6]{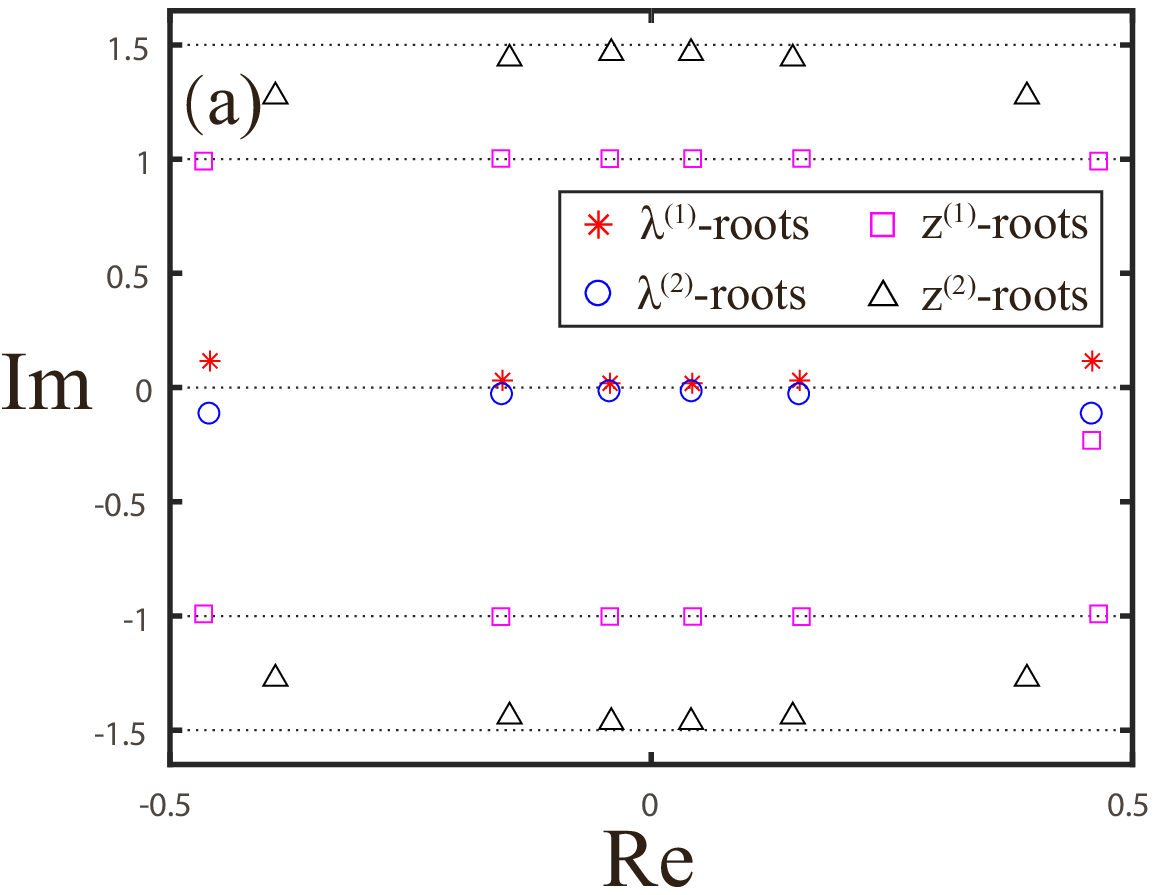}}
  \subfigure{\label{fig3:subfig:b} 
    \includegraphics[scale=0.6]{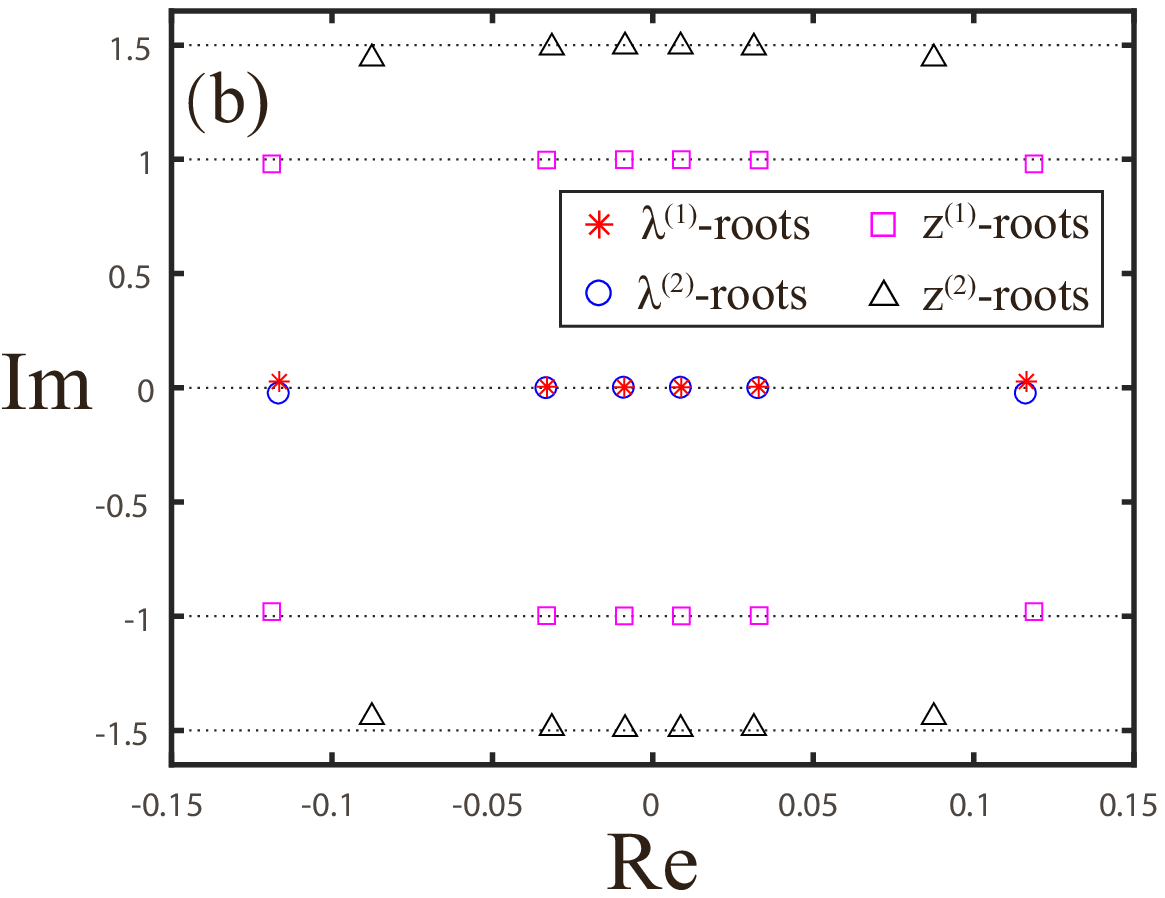}
    }
  \caption{(Colour online) The distributions of the $\lambda^{(1)}$-roots (red $\ast$), $\lambda^{(2)}$-roots (blue $\circ$), $z^{(1)}$-roots (mauve $\square$) and $z^{(2)}$-roots (black $\vartriangle$) for the state with the maximus $\Lambda^{(Q)}(0)$ at finite temperatures $T=1$ (a) and $T=5$ (b). In both plots, we set $N=12$.}\label{fig3}
\end{figure}

Similarly, the normalized eigenvalues $\bar{\Lambda}_1^{(Q)}(u)$ and $\bar{\Lambda}_2^{(Q)}(u)$ are defined
\bea
&&\bar{\Lambda}_1^{(Q)}(u)=\frac{\Lambda_1^{(Q)}(u)}{(u-\eta\tau+\eta)^M(u+\eta\tau-\eta)^M},\label{Normalize-su3-1}\\[4pt]
&&\bar{\Lambda}_2^{(Q)}(u)=\frac{\Lambda_2^{(Q)}(u)}{(u-\eta\tau+\frac{3}{2}\eta)^M(u+\eta\tau-\frac{3}{2}\eta)^M}.\label{Normalize-su3-2}
\eea
\begin{figure}[htbp]
  \centering
  \subfigure{\label{fig4:subfig:a} 
    \includegraphics[scale=0.6]{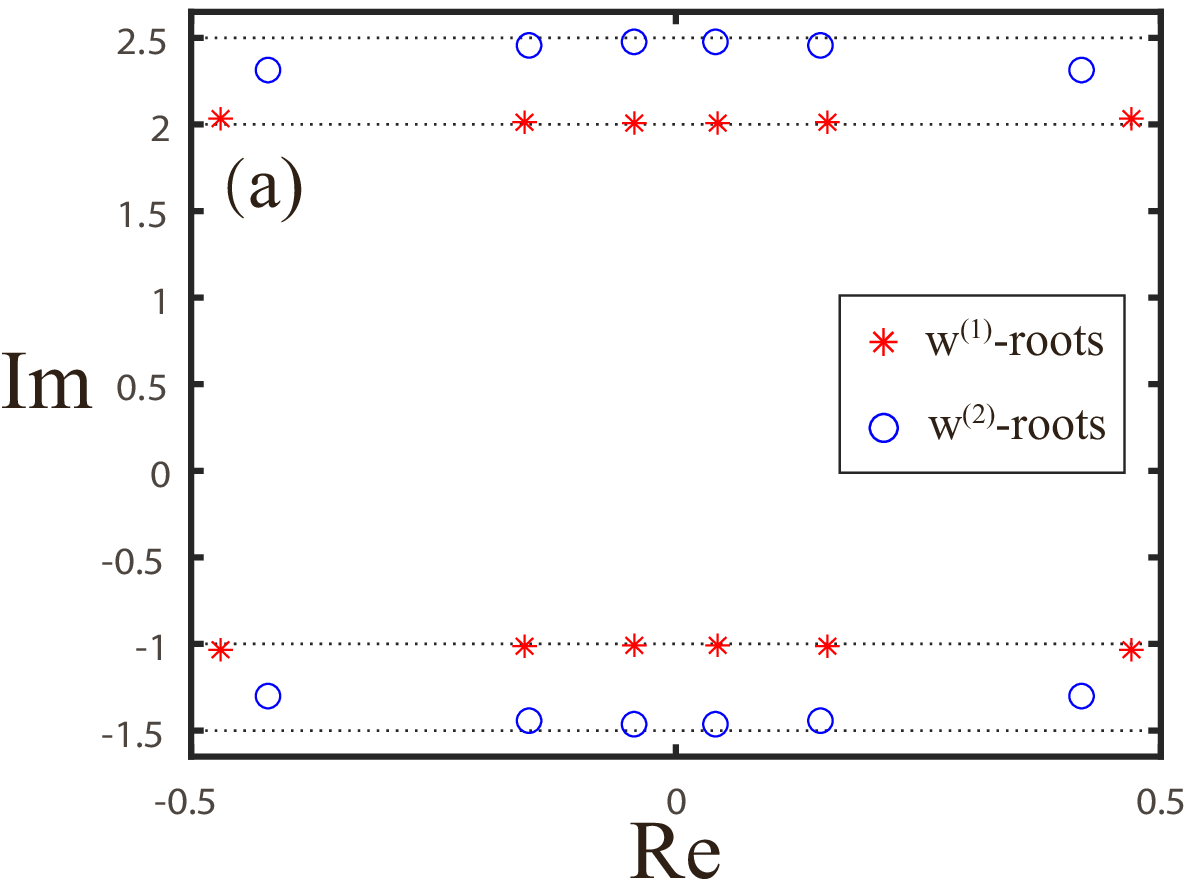}}
  \subfigure{\label{fig4:subfig:b} 
    \includegraphics[scale=0.6]{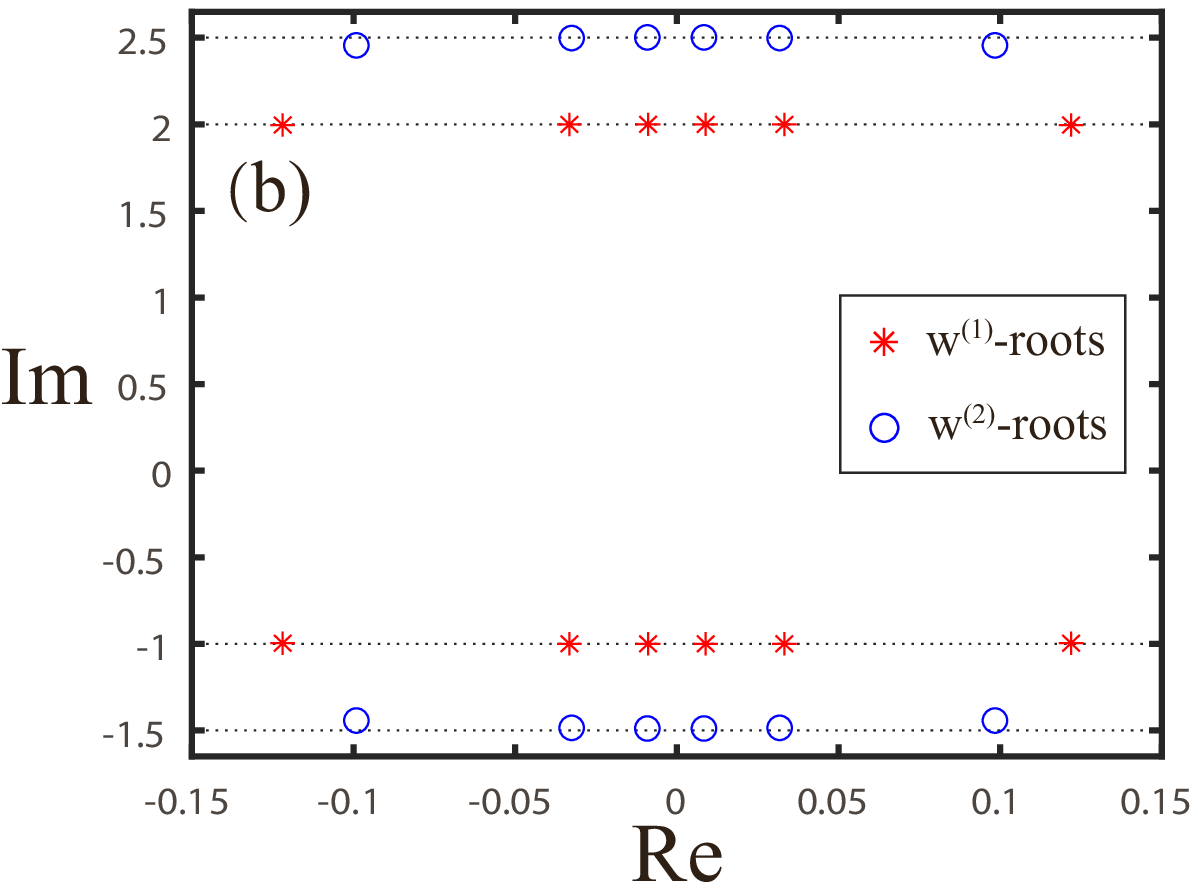}
    }
  \caption{(Colour online) The distributions of the $\omega^{(1)}$-roots (red $*$), and $\omega^{(2)}$-roots (blue $\circ$) for the state with the maximus $\Lambda^{(Q)}(0)$ at finite temperatures $T=1$ (a) and $T=5$ (b). In both plots, we set $N=12$.}\label{fig4}
\end{figure}
\hspace{-0.1cm}Numerical study with some small $N$ (up to 12) for the distributions in figure \ref{fig3} of the roots of $\{\Lambda_i^{(Q)}(u)|i=1,2\}$ for the state with $|\Lambda_1^{(Q)}(0)|_{max}$  indicts that the corresponding $\{\bar{\Lambda}_i^{(Q)}(u)|i=1,2\}$ have the decompositions
\bea
&&\bar{\Lambda}_1^{(Q)}(u)=b(\beta)\,\frac{\prod_{j=1}^M(u-u^{(+)}_{1,j}-\eta)\,(u-u^{(-)}_{1,j}+\eta)}{(u+\eta\tau-\eta)^M\,(u-\eta\tau+\eta)^M},\label{Zero-points-Z-su31}\\[4pt]
&&\bar{\Lambda}_2^{(Q)}(u)=b(\beta)\,\frac{\prod_{j=1}^M(u-u^{(+)}_{2,j}-\frac{3}{2}\eta)\,(u-u^{(-)}_{2,j}+\frac{3}{2}\eta)}{(u+\eta\tau-\frac{3}{2}\eta)^M\,(u-\eta\tau+\frac{3}{2}\eta)^M},
\label{Zero-points-Z-su32}
\eea
where $\{{\rm Im}(u^{(\pm)}_{i,j})\sim 0|i=1,2\}$ for a large $N$. The Bethe ansatz solutions (given in Appendix D, see the $T-Q$ relations below (\ref{T-Q-su3-1}) and (\ref{T-Q-su3-2}) ) indeed  confirm that the roots of $\{\Lambda_i^{(Q)}(u)|i=1,2\}$ for the state with  $|\Lambda^{(Q)}(0)|_{max}$ do have the distributions (\ref{Zero-points-Z-su31}) and (\ref{Zero-points-Z-su32}).
With the help of the $t-W$ relations (\ref{Eigen-id-su31})-(\ref{Eigen-id-su32}) and numerical results as shown in figure \ref{fig4}, we can derive that the eigenvalues $\{W_i^{(Q)}(u)|i=1,2\}$ for the state with  $|\Lambda_1^{(Q)}(0)|_{max}$ have the decompositions
\bea
&&W_1^{(Q)}(u)=(b^2(\beta)-b(\beta))\,\prod_{j=1}^M(u-w^{(+)}_{1,j}-2\eta)(u-w^{(-)}_{1,j}+\eta), \label{Zero-points-W-su31}\\[4pt]
&&W_2^{(Q)}(u)=(b^2(\beta)-b(\beta))\,\prod_{j=1}^M(u-w^{(+)}_{2,j}-\frac{5}{2}\eta)(u-w^{(-)}_{2,j}+\frac{3}{2}\eta), \label{Zero-points-W-su32}
\eea
where $\{{\rm Im}(w^{(\pm)}_{i,j})\sim 0|i=1,2\}$ for a large $N$. Then the $t-W$ relations (\ref{Eigen-id-su31}) and (\ref{Eigen-id-su32}) for the state with  $|\Lambda_1^{(Q)}(0)|_{max}$ become
\bea
\hspace{-0.88truecm}&&\hspace{-0.88truecm}\quad\quad\bar{\Lambda}_1^{(Q)}(u\hspace{-0.02truecm}+\hspace{-0.02truecm}\frac{\eta}{2})
 \bar{\Lambda}_1^{(Q)}(u\hspace{-0.02truecm}-\hspace{-0.02truecm}\frac{\eta}{2})\no\\[8pt]
\hspace{-0.48truecm}&&\hspace{-0.48truecm}\qquad\qquad= b^2(\beta)\frac{\prod_{j=1}^M\hspace{-0.02truecm}(u\hspace{-0.02truecm}-\hspace{-0.02truecm}u^{(-)}_{1,j}\hspace{-0.02truecm}+\hspace{-0.02truecm}\frac{3}{2}\eta)
 (u\hspace{-0.02truecm}-\hspace{-0.02truecm}u^{(+)}_{1,j}\hspace{-0.02truecm}-\hspace{-0.02truecm}\frac{\eta}{2})
(u\hspace{-0.02truecm}-\hspace{-0.02truecm}u^{(-)}_{1,j}\hspace{-0.02truecm}+\hspace{-0.02truecm}\frac{\eta}{2})
(u\hspace{-0.02truecm}-\hspace{-0.02truecm}u^{(+)}_{1,j}\hspace{-0.02truecm}-\hspace{-0.02truecm}\frac{3}{2}\eta)}
{(u-\eta\tau+\frac{3}{2}\eta)^M (u+\eta\tau-\frac{\eta}{2})^M (u-\eta\tau+\frac{\eta}{2})^M (u+\eta\tau-\frac{3}{2}\eta)^M}\no\\[8pt]
\hspace{-0.48truecm}&&\hspace{-0.48truecm}\qquad\qquad=b(\beta)\frac{(u\hspace{-0.02truecm}+\hspace{-0.02truecm}\eta\tau\hspace{-0.02truecm}+\hspace{-0.02truecm}\frac{\eta}{2})^M
(u\hspace{-0.02truecm}-\hspace{-0.02truecm}\eta\tau\hspace{-0.02truecm}-\hspace{-0.02truecm}\frac{\eta}{2})^M}
{(u\hspace{-0.02truecm}-\hspace{-0.02truecm}\eta\tau\hspace{-0.02truecm}+\hspace{-0.02truecm}\frac{\eta}{2})^M
(u\hspace{-0.02truecm}+\hspace{-0.02truecm}\eta\tau\hspace{-0.02truecm}-\hspace{-0.02truecm}
\frac{\eta}{2})^M}\frac{\prod_{j=1}^M(u\hspace{-0.02truecm}-\hspace{-0.02truecm}u^{(+)}_{2,j}\hspace{-0.02truecm}-\hspace{-0.02truecm}\frac{3}{2}\eta)
(u\hspace{-0.02truecm}-\hspace{-0.02truecm}u^{(-)}_{2,j}\hspace{-0.02truecm}+\hspace{-0.02truecm}\frac{3}{2}\eta)}
{(u\hspace{-0.02truecm}+\hspace{-0.02truecm}\eta\tau\hspace{-0.02truecm}-\hspace{-0.02truecm}\frac{3}{2}\eta)^M
(u\hspace{-0.02truecm}-\hspace{-0.02truecm}\eta\tau\hspace{-0.02truecm}+\hspace{-0.02truecm}\frac{3}{2}\eta)^M}\no\\[8pt]
&&\qquad\qquad\qquad\qquad+
(b^2(\beta)-b(\beta))
\frac{\prod_{j=1}^M(u\hspace{-0.02truecm}-\hspace{-0.02truecm}w^{(+)}_{1,j}\hspace{-0.02truecm}-\hspace{-0.02truecm}\frac{3}{2}\eta)
(u\hspace{-0.02truecm}-\hspace{-0.02truecm}w^{(-)}_{1,j}\hspace{-0.02truecm}+\hspace{-0.02truecm}\frac{3}{2}\eta)}
{(u\hspace{-0.02truecm}+\hspace{-0.02truecm}\eta\tau\hspace{-0.02truecm}-\hspace{-0.02truecm}\frac{3}{2}\eta)^M
(u\hspace{-0.02truecm}-\hspace{-0.02truecm}\eta\tau\hspace{-0.02truecm}+\hspace{-0.02truecm}\frac{3}{2}\eta)^M}\no\\[8pt]
\hspace{-0.48truecm}&&\hspace{-0.48truecm}\qquad\qquad=b(\beta)q(u)\bar{\lambda}_2(u)
+(b^2(\beta)-b(\beta))\bar{w}_1(u)+O(\frac{1}{N}),\label{t-W-relation-su3-1}
\eea
\bea
\hspace{-0.88truecm}&&\hspace{-0.88truecm}\quad\quad\bar{\Lambda}_2^{(Q)}(u\hspace{-0.02truecm}+\hspace{-0.02truecm}\frac{\eta}{2})
 \bar{\Lambda}_2^{(Q)}(u\hspace{-0.02truecm}-\hspace{-0.02truecm}\frac{\eta}{2})\no\\[8pt]
\hspace{-0.48truecm}&&\hspace{-0.48truecm}\qquad\qquad= b^2(\beta)\frac{\prod_{j=1}^M\hspace{-0.02truecm}(u\hspace{-0.02truecm}-\hspace{-0.02truecm}u^{(-)}_{2,j}\hspace{-0.02truecm}+\hspace{-0.02truecm}2\eta)
 (u\hspace{-0.02truecm}-\hspace{-0.02truecm}u^{(+)}_{2,j}\hspace{-0.02truecm}-\hspace{-0.02truecm}\eta)
(u\hspace{-0.02truecm}-\hspace{-0.02truecm}u^{(-)}_{2,j}\hspace{-0.02truecm}+\hspace{-0.02truecm}\eta)
(u\hspace{-0.02truecm}-\hspace{-0.02truecm}u^{(+)}_{2,j}\hspace{-0.02truecm}-\hspace{-0.02truecm}2\eta)}
{(u-\eta\tau+2\eta)^M (u+\eta\tau-\eta)^M (u-\eta\tau+\eta)^M (u+\eta\tau-2\eta)^M}\no\\[8pt]
\hspace{-0.48truecm}&&\hspace{-0.48truecm}\qquad\qquad=b(\beta)\frac{\prod_{j=1}^M(u\hspace{-0.02truecm}-\hspace{-0.02truecm}u^{(+)}_{1,j}\hspace{-0.02truecm}-\hspace{-0.02truecm}\eta)
(u\hspace{-0.02truecm}-\hspace{-0.02truecm}u^{(-)}_{1,j}\hspace{-0.02truecm}+\hspace{-0.02truecm}\eta)}
{(u\hspace{-0.02truecm}+\hspace{-0.02truecm}\eta\tau\hspace{-0.02truecm}-\hspace{-0.02truecm}\eta)^M
(u\hspace{-0.02truecm}-\hspace{-0.02truecm}\eta\tau\hspace{-0.02truecm}+\hspace{-0.02truecm}\eta)^M}\no\\[8pt]
&&\qquad\qquad\qquad\qquad+(b^2(\beta)-b(\beta))
\frac{\prod_{j=1}^M(u\hspace{-0.02truecm}-\hspace{-0.02truecm}w^{(+)}_{2,j}\hspace{-0.02truecm}-\hspace{-0.02truecm}2\eta)
(u\hspace{-0.02truecm}-\hspace{-0.02truecm}w^{(-)}_{2,j}\hspace{-0.02truecm}+\hspace{-0.02truecm}2\eta)}
{(u\hspace{-0.02truecm}+\hspace{-0.02truecm}\eta\tau\hspace{-0.02truecm}-\hspace{-0.02truecm}2\eta)^M
(u\hspace{-0.02truecm}-\hspace{-0.02truecm}\eta\tau\hspace{-0.02truecm}+\hspace{-0.02truecm}2\eta)^M}\no\\[8pt]
\hspace{-0.48truecm}&&\hspace{-0.48truecm}\qquad\qquad=b(\beta)\bar{\lambda}_1(u)+(b^2(\beta)-b(\beta))\bar{w}_2(u)+O(\frac{1}{N}),\label{t-W-relation-su3-2}
\eea
where the functions $\bar{\lambda}_1(u)$, $\bar{\lambda}_2(u)$, $\bar{w}_1(u)$ and $\bar{w}_2(u)$  are
\bea
\bar{\lambda}_1(u)\hspace{-0.2truecm}&=&\hspace{-0.2truecm}\lim_{M\rightarrow \infty}\,
\frac{\prod_{j=1}^M(u\hspace{-0.02truecm}-\hspace{-0.02truecm}u^{(+)}_{1,j}\hspace{-0.02truecm}-\hspace{-0.02truecm}\eta)
(u\hspace{-0.02truecm}-\hspace{-0.02truecm}u^{(-)}_{1,j}\hspace{-0.02truecm}+\hspace{-0.02truecm}\eta)}
{(u\hspace{-0.02truecm}+\hspace{-0.02truecm}\eta\tau\hspace{-0.02truecm}-\hspace{-0.02truecm}\eta)^M
(u\hspace{-0.02truecm}-\hspace{-0.02truecm}\eta\tau\hspace{-0.02truecm}+\hspace{-0.02truecm}\eta)^M}
\stackrel{{\rm def}}{=}e^{-\b\, \bar{\varepsilon}_{1}(u)},\label{l1-function}\\[8pt]
\bar{\lambda}_2(u)\hspace{-0.2truecm}&=&\hspace{-0.2truecm}\lim_{M\rightarrow \infty}\,
\frac{\prod_{j=1}^M(u\hspace{-0.02truecm}-\hspace{-0.02truecm}u^{(+)}_{2,j}\hspace{-0.02truecm}-\hspace{-0.02truecm}\frac{3}{2}\eta)
(u\hspace{-0.02truecm}-\hspace{-0.02truecm}u^{(-)}_{2,j}\hspace{-0.02truecm}+\hspace{-0.02truecm}\frac{3}{2}\eta)}
{(u\hspace{-0.02truecm}+\hspace{-0.02truecm}\eta\tau\hspace{-0.02truecm}-\hspace{-0.02truecm}\frac{3}{2}\eta)^M
(u\hspace{-0.02truecm}-\hspace{-0.02truecm}\eta\tau\hspace{-0.02truecm}+\hspace{-0.02truecm}\frac{3}{2}\eta)^M}
\stackrel{{\rm def}}{=}e^{-\b\, \bar{\varepsilon}_{2}(u)},\label{l2-function}\\[8pt]
\bar{w}_1(u)\hspace{-0.2truecm}&=&\hspace{-0.2truecm}\lim_{M\rightarrow \infty}\,
\frac{\prod_{j=1}^M(u\hspace{-0.02truecm}-\hspace{-0.02truecm}w^{(+)}_{1,j}\hspace{-0.02truecm}-\hspace{-0.02truecm}\frac{3}{2}\eta)
(u\hspace{-0.02truecm}-\hspace{-0.02truecm}w^{(-)}_{1,j}\hspace{-0.02truecm}+\hspace{-0.02truecm}\frac{3}{2}\eta)}
{(u\hspace{-0.02truecm}+\hspace{-0.02truecm}\eta\tau\hspace{-0.02truecm}-\hspace{-0.02truecm}\frac{3}{2}\eta)^M
(u\hspace{-0.02truecm}-\hspace{-0.02truecm}\eta\tau\hspace{-0.02truecm}+\hspace{-0.02truecm}\frac{3}{2}\eta)^M}
\stackrel{{\rm def}}{=}e^{-\b\, \bar{\varepsilon}_{3}(u)},\label{w1-function}\\[8pt]
\bar{w}_2(u)\hspace{-0.2truecm}&=&\hspace{-0.2truecm}\lim_{M\rightarrow \infty}\,
\frac{\prod_{j=1}^M(u\hspace{-0.02truecm}-\hspace{-0.02truecm}w^{(+)}_{2,j}\hspace{-0.02truecm}-\hspace{-0.02truecm}2\eta)
(u\hspace{-0.02truecm}-\hspace{-0.02truecm}w^{(-)}_{2,j}\hspace{-0.02truecm}+\hspace{-0.02truecm}2\eta)}
{(u\hspace{-0.02truecm}+\hspace{-0.02truecm}\eta\tau\hspace{-0.02truecm}-\hspace{-0.02truecm}2\eta)^M
(u\hspace{-0.02truecm}-\hspace{-0.02truecm}\eta\tau\hspace{-0.02truecm}+\hspace{-0.02truecm}2\eta)^M}
\stackrel{{\rm def}}{=}e^{-\b\, \bar{\varepsilon}_{4}(u)}.\label{w2-function}
\eea
It is remarked that the functions $\{\bar{\varepsilon}_i(u)|i=1,...,4\}$ satisfy the analytic properties:
\bea
&&\bar{\varepsilon}_1(u) {\rm ~is~ analytic~ except~ some~ singularities~ on~ the~ axis}~ {\rm Im}(u)=\pm 1,\no\\
&&\bar{\varepsilon}_2(u)\,\,{\rm and}\,\, \bar{\varepsilon}_3(u) {\rm ~are~ analytic~ except~ some~ singularities~ on~ the~ axis}~ {\rm Im}(u)=\pm \frac{3}{2},\no\\
&&\bar{\varepsilon}_4(u) {\rm ~is~ analytic~ except~ some~ singularities~ on~ the~ axis}~ {\rm Im}(u)=\pm 2,\no\\
&&\lim_{u\rightarrow \infty}\bar{\varepsilon}_i(u)=0,\quad i=1,...,4.\label{Property-epsion-function-su3}
\eea
\subsection{Integral representations and free energy of the periodic $SU(3)$ model}
\label{su3-2}
The decompositions (\ref{Zero-points-Z-su31})-(\ref{Zero-points-Z-su32}) and the very $t-W$ relations (\ref{t-W-relation-su3-1})-(\ref{t-W-relation-su3-2}) allow us to give the integral representations of $\bar{\Lambda}_1^{(Q)}(u)$ and $\bar{\Lambda}_2^{(Q)}(u)$
\bea
\ln \bar{\Lambda}_1^{(Q)}(u)\hspace{-0.2truecm}&=&\hspace{-0.2truecm}\ln b(\beta)+\frac{1}{2\pi i}\oint _{\mathcal{C}_1^{\prime}}dv \,\frac{\ln\lt((q(v)e^{-\b \bar{\varepsilon}_2(v)}+(b(\beta)-1)
e^{-\b \bar{\varepsilon}_3(v)})/b(\beta)\rt)}{u-v-\frac{\eta}{2}}\no\\[8pt]
&&\qquad\quad\quad\quad +\frac{1}{2\pi i}\oint _{\mathcal{C}^{\prime}_2}dv \,\frac{\ln\lt((q(v)e^{-\b \bar{\varepsilon}_2(v)}+(b(\beta)-1)
e^{-\b \bar{\varepsilon}_3(v)})/b(\beta)\rt)}{u-v+\frac{\eta}{2}},
\label{Integral-rep-su31}
\eea
\bea
\hspace{-0.6truecm}\ln \bar{\Lambda}_2^{(Q)}(u)\hspace{-0.2truecm}&=&\hspace{-0.2truecm}\ln b(\beta)+\frac{1}{2\pi i}\oint _{\mathcal{C}_3^{\prime}}dv \,\frac{\ln\lt(e^{-\b \bar{\varepsilon}_1(v)}+(b(\beta)-1)
e^{-\b \bar{\varepsilon}_4(v)})/b(\beta)\rt)}{u-v-\frac{\eta}{2}}\no\\[8pt]
&&\qquad\quad\quad\quad +\frac{1}{2\pi i}\oint _{\mathcal{C}^{\prime}_4}dv \,\frac{\ln\lt((e^{-\b \bar{\varepsilon}_1(v)}+(b(\beta)-1)
e^{-\b \bar{\varepsilon}_4(v)})/b(\beta)\rt)}{u-v+\frac{\eta}{2}},
\label{Integral-rep-su32}
\eea
where the closed integral contour $\mathcal{C}_1^{\prime}$ (or $\mathcal{C}_2^{\prime}$) is surrounding the axis of ${\rm Im}(v)\hspace{-0.085truecm}=\hspace{-0.085truecm}\frac{1}{2}\,({\rm or} -\frac{1}{2})$, while $\mathcal{C}_3^{\prime}$ ( or $\mathcal{C}_4^{\prime}$) is surrounding the axis of ${\rm Im}(v)\hspace{-0.085truecm}=\hspace{-0.085truecm}1\,({\rm or} -1)$.
With the help of the $t-W$ relations (\ref{t-W-relation-su3-1})-(\ref{t-W-relation-su3-2}) and the integral representations (\ref{Integral-rep-su31})-(\ref{Integral-rep-su32}), we can derive two NLIEs of the functions $\{\bar{\varepsilon}_i(u)|i=1,...,4\}$
\bea
\hspace{-0.8truecm}&&\hspace{-0.8truecm}\ln(q(u)e^{-\b \bar{\varepsilon}_2(u)}\hspace{-0.02truecm}+\hspace{-0.02truecm}(b(\beta)\hspace{-0.02truecm}-\hspace{-0.02truecm}1)
e^{-\b \bar{\varepsilon}_3(u)})=\ln b(\beta)\no\\[8pt]
\hspace{-1.2truecm}&&+\frac{1}{2\pi i}\hspace{-0.02truecm}\oint _{\mathcal{C}^{\prime}_1}\hspace{-0.02truecm}dv \hspace{-0.02truecm}(\frac{1}{u\hspace{-0.02truecm}-\hspace{-0.02truecm}v}
\hspace{-0.02truecm}+\hspace{-0.02truecm}\frac{1}{u\hspace{-0.02truecm}-\hspace{-0.02truecm}v\hspace{-0.02truecm}-\hspace{-0.02truecm}\eta})
\hspace{-0.02truecm}\ln\lt((q(v)e^{-\b \bar{\varepsilon}_2(v)}\hspace{-0.02truecm}+\hspace{-0.02truecm}(b(\beta)\hspace{-0.02truecm}-\hspace{-0.02truecm}1)
e^{-\b \bar{\varepsilon}_3(v)})/b(\beta)\rt)\no\\[8pt]
\hspace{-1.2truecm}&&+\frac{1}{2\pi i}\hspace{-0.02truecm}\oint _{\mathcal{C}^{\prime}_2}\hspace{-0.02truecm}dv \hspace{-0.02truecm}(\frac{1}{u\hspace{-0.02truecm}-\hspace{-0.02truecm}v\hspace{-0.02truecm}+\hspace{-0.02truecm}\eta}
\hspace{-0.02truecm}+\hspace{-0.02truecm}\frac{1}{u\hspace{-0.02truecm}-\hspace{-0.02truecm}v})
\hspace{-0.02truecm}\ln\lt((q(v)e^{-\b \bar{\varepsilon}_2(v)}\hspace{-0.02truecm}+\hspace{-0.02truecm}(b(\beta)\hspace{-0.02truecm}-\hspace{-0.02truecm}1)
e^{-\b \bar{\varepsilon}_3(v)})/b(\beta)\rt),\label{Integral-Main-su31}
\eea
\bea
\hspace{-2truecm}&&\hspace{-0.9truecm}\ln(e^{-\b \bar{\varepsilon}_1(u)}\hspace{-0.02truecm}+\hspace{-0.02truecm}(b(\beta)\hspace{-0.02truecm}-\hspace{-0.02truecm}1)
e^{-\b \bar{\varepsilon}_4(u)})=\ln b(\beta)\no\\[8pt]
\hspace{-1.2truecm}&&+\frac{1}{2\pi i}\hspace{-0.02truecm}\oint _{\mathcal{C}^{\prime}_3}\hspace{-0.02truecm}dv \hspace{-0.02truecm}(\frac{1}{u\hspace{-0.02truecm}-\hspace{-0.02truecm}v}
\hspace{-0.02truecm}+\hspace{-0.02truecm}\frac{1}{u\hspace{-0.02truecm}-\hspace{-0.02truecm}v\hspace{-0.02truecm}-\hspace{-0.02truecm}\eta})
\hspace{-0.02truecm}\ln\lt((e^{-\b \bar{\varepsilon}_1(v)}\hspace{-0.02truecm}+\hspace{-0.02truecm}(b(\beta)\hspace{-0.02truecm}-\hspace{-0.02truecm}1)
e^{-\b \bar{\varepsilon}_4(v)})/b(\beta)\rt)\no\\[8pt]
\hspace{-1.2truecm}&&+\frac{1}{2\pi i}\hspace{-0.02truecm}\oint _{\mathcal{C}^{\prime}_4}\hspace{-0.02truecm}dv \hspace{-0.02truecm}(\frac{1}{u\hspace{-0.02truecm}-\hspace{-0.02truecm}v\hspace{-0.02truecm}+\hspace{-0.02truecm}\eta}
\hspace{-0.02truecm}+\hspace{-0.02truecm}\frac{1}{u\hspace{-0.02truecm}-\hspace{-0.02truecm}v})
\hspace{-0.02truecm}\ln\lt((e^{-\b \bar{\varepsilon}_1(v)}\hspace{-0.02truecm}+\hspace{-0.02truecm}(b(\beta)\hspace{-0.02truecm}-\hspace{-0.02truecm}1)
e^{-\b \bar{\varepsilon}_4(v)})/b(\beta)\rt).\label{Integral-Main-su32}
\eea
It is believed that the analytic properties (\ref{Property-epsion-function-su3}), the integral representations (\ref{Integral-rep-su31})-(\ref{Integral-rep-su32}) and the NLIEs (\ref{Integral-Main-su31})-(\ref{Integral-Main-su32}) might completely determine the functions $\{\bar{\varepsilon}_i(u)|=1,...,4\}$.
\begin{figure}[htbp]
\centering
\includegraphics[scale=0.6]{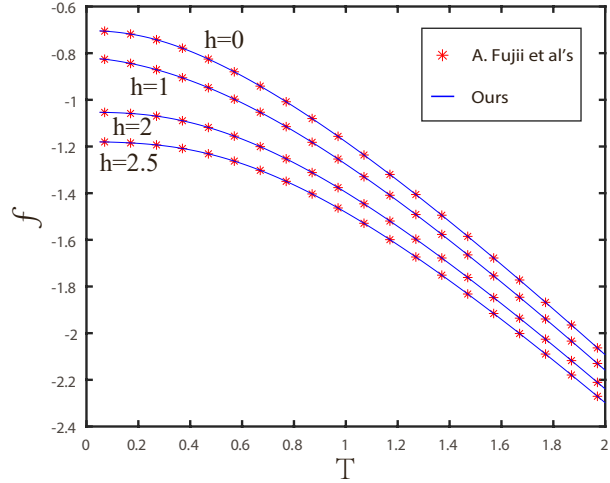}
  \caption{Free energy $f$ vs $T$ for the periodic $SU(3)$ chain in different magnetic fields.}\label{fig5}
\end{figure}
Due to the fact that the roots and the poles of $ \bar{\Lambda}_1^{(Q)}(u)$ locate nearly on the two lines with imaginary parts close to $\pm 1$ (see the decomposition (\ref{Zero-points-Z-su31})),
we can use the Fourier transformation to obtain another integral representation of $\bar{\Lambda}_1^{(Q)}(u)$
\bea
\ln \bar{\Lambda}_1^{(Q)}(u)\hspace{-0.2truecm}&=&\hspace{-0.2truecm} \int_{-\infty}^{+\infty}\frac{dv}{2\cosh\pi(u-v)}\lt\{\ln\bar{\Lambda}_1^{(Q)}(v+\frac{\eta}{2})+
\ln\bar{\Lambda}_1^{(Q)}(v-\frac{\eta}{2}) \rt\}\no\\[8pt]
\hspace{-0.2truecm}&=&\hspace{-0.2truecm}\int_{-\infty}^{+\infty}\frac{dv}{2\cosh\pi(u-v)}\lt\{b(\beta)q(v)e^{-\b \bar{\varepsilon}_2(v)}\hspace{-0.02truecm}+\hspace{-0.02truecm}(b^2(\beta)\hspace{-0.02truecm}-\hspace{-0.02truecm}b(\beta))
e^{-\b \bar{\varepsilon}_3(v)}  \rt\}.\no
\eea
Finally we can obtain the free energy of the periodic $SU(3)$ chain described by the
Hamiltonian (\ref{su3}) with an external field $h$ as
\bea
f(\beta)\hspace{-0.2truecm}&=&\hspace{-0.2truecm}2J-\frac{1}{\b}\ln \bar{\Lambda}_1^{(Q)}(0)\no\\[8pt]
\hspace{-0.2truecm}&=&\hspace{-0.2truecm}2J-\frac{1}{\b} \int_{-\infty}^{+\infty}\frac{dv}
{2\cosh\pi v}
\ln \lt(b(\beta)q(v)e^{-\b \bar{\varepsilon}_2(v)}+(b^2(\beta)-b(\beta))
e^{-\b \bar{\varepsilon}_3(v)}\rt).\label{free-energy-su3}
\eea
Using the numerical iterative procedure in Appendix E, we obtain the free energy $f$ variation with temperature $T$ in different magnetic fields as shown in figure \ref{fig5}. From this figure, we can conclude that our result coincides well that of \cite{FUJII1999} with a different approach. Moreover, we can obtain  the HTE of the free energy of the $SU(3)$-invariant spin chain described by the Hamiltonian (\ref{su3})-(\ref{BC-su3}) (the details of the derivation is given in Appendix F)
\bea
f/T=-\ln(2\cosh(\frac{h}{2T})+1)+\frac{J}{T}\frac{4}{2\cosh(\frac{h}{2T})+1}
-\frac{J^2}{T^2}\frac{24\cosh(\frac{h}{2T})}{(2\cosh(\frac{h}{2T})+1)^2}+\cdot\cdot\cdot.\label{HTE-su(3)}
\eea

\section{Conclusions}
\label{Con}

In this paper, we have studied the thermodynamics of the Heisenberg  chain at a finite temperature in anti-ferromagnetic regime via the recent developed $t-W$ method \cite{PhysRevB.102.085115,Qia21}. A novel nonlinear integral equation (\ref{Integral-Main}) which involves only one auxiliary function $\epsilon(u)$ has been given via  the  $t-W$ relation (\ref{t-W-relation-op}) satisfied by the associated transfer matrices. Together with some analytic property of the function $\epsilon(u)$ (see (\ref{Property-epsion-function-1}) and (\ref{epsilon-function})), we solve the NLIE and obtain the free energy of the Heisenberg chain in different magnetic fields. Our results coincide well with those obtained by the other methods. Moreover, using the fusion technique we have obtained the $t-W$ relations (\ref{t-W-su(n)})-(\ref{t-W-su(n)-1}) among the transfer matrices for the quantum integrable systems associated with $SU(n)$, which allow one to derive the associated NLIEs. Taking the $SU(3)$-invariant quantum chain as an example, we construct the corresponding NLIEs which involve only four auxiliary functions. Solving the NLIEs, we  obtain the free energy of the model.

We have proposed a more direct, efficient and easily extensible procedure to construct the associated  NLIEs of obtaining  free energy of the quantum spin chains.  Our method can be easily generalized to the quantum integrable models solved by the off diagonal Bethe ansatz method \cite{Li19,Li19-1,Li19-2}, which are associated with other Lie algebras such as $B_n$, $C_n$ and $D_n$. Moreover, for the $SU(n)$-case, only $2n-2$ functions have been involved in  to obtain the eigenvalue $\Lambda^{(Q)}(u)$  of the original quantum transfer matrix.

\section*{Acknowledgments}

The financial supports from the National Program for Basic
Research of MOST (Grant Nos. 2016 YFA0300600 and 2016YFA0302104),
National Natural Science Foundation of China (Grant Nos. 12074410, 12047502, 11934015,
11975183, 11947301 and 11774397), Major Basic Research Program of Natural Science of
Shaanxi Province (Grant Nos. 2017KCT-12 and 2017ZDJC-32), Australian
Research Council (Grant No. DP 190101529), the Strategic
Priority Research Program of the Chinese Academy of Sciences (Grant No. XDB33000000),
the fellowship of China Postdoctoral Science Foundation (Grant No. 2020M680724), and
Double First-Class University Construction Project of
Northwest University are gratefully acknowledged.


\section*{Appendix A: Proof of the $t-W$ relations (\ref{t-W-relation-op}) and (\ref{t-W-su(n)}) }
\setcounter{equation}{0}
\renewcommand{\theequation}{A.\arabic{equation}}
In this appendix we shall prove the  operator relation (\ref{t-W-relation-op}) among the transfer matrices by using
the fusion technique \cite{Kul81,Kir86} of the $R$-matrix.

For this purpose, let us introduce the (anti)symmetric subspaces $W^{(\pm)}$ of $\mathbb{C}^2\otimes \mathbb{C}^2$: $W^{(\pm)}=P^{(\pm)}_{12}\mathbb{C}^2\otimes \mathbb{C}^2$. Let $\{|i\rangle |i=1,2\}$ be an orthnormal basis of $\mathbb{C}^2$.
It is easy to see that $W^{(+)}$ is a $3$-dimensional subspace spanned by the orthnormal basis $\{|11\rangle,\,\frac{1}{\sqrt{2}}(|12\rangle+|21\rangle),\,|22\rangle\}$, while $W^{(-)}$ is an $1$-dimensional subspace spanned by
$\{\frac{1}{\sqrt{2}}(|12\rangle-|21\rangle)\}$.  The operator $\s^z_1+\s^z_2$ acts on $W^{(\pm)}$ invariantly respectively. Denoted
the action of $\s^z_1+\s^z_2$ on the subspace $W^{(+)}$ by $\s^z_{\{12\}}$, it becomes a $3\times 3$-matrix.  In the basis $\{|11\rangle,\,\frac{1}{\sqrt{2}}(|12\rangle+|21\rangle),\,|22\rangle\}$, it reads
\bea
\s^z_{\{12\}}=\lt(\begin{array}{ccc}2&&\\
&0&\\&&-2\end{array}\rt).
\eea
The QYBE (\ref{QYB}) and the fusion condition (\ref{Fusion-con}) allow us to derive the relation
\bea
R_{23}(u)R_{13}(u\hspace{-0.08truecm}-\hspace{-0.08truecm}\eta)P^{(-)}_{12}
=P^{(-)}_{12}R_{23}(u)R_{13}(u\hspace{-0.08truecm}-\hspace{-0.08truecm}\eta)P^{(-)}_{12}
=(u+\eta)(u-\eta)\times {\rm id}.\label{Q-determinant}
\eea
Direct calculation shows that
\bea
P^{(+)}_{12}\,R_{23}(u)\,R_{13}(u-\eta)\,P^{(+)}_{12}
=u\, R^{(1,\frac{1}{2})}_{\{12\}\,3}(u),\label{Fusion-R}
\eea
where the fused $R$-matrix $R^{(1,\frac{1}{2})}_{\{12\}\,3}(u)$, in the basis $\{|11\rangle,\,\frac{1}{\sqrt{2}}(|12\rangle+|21\rangle),\,|22\rangle\}$, reads
\bea
R^{(1,\frac{1}{2})}_{\{12\}\,3}(u)=\lt(\begin{array}{cccccc}
u+\eta&&&&&\\[6pt]
&u-\eta&\sqrt{2}&&&\\[6pt]
&\sqrt{2}&u&&&\\[6pt]
&&& u&\sqrt{2}&\\[6pt]
&&&\sqrt{2}&u-\eta&\\[6pt]
&&&&&u+\eta
\end{array}\rt).\label{Fusion-R-1}
\eea

Keeping (\ref{trans-per}) in mind, let us introduce one-row monodromy matrix
\bea
T_0(u)=R_{0N}(u-\theta_N)\cdots R_{01}(u-\theta_1).
\eea
The relations (\ref{Q-determinant})  and (\ref{Fusion-R}) lead to
\bea
&&P^{(-)}_{12}T_2(u)\,T_1(u-\eta)P^{(-)}_{12}=a(u)\,d(u-\eta)\times {\rm id},\label{Q-Det-1-1}\\[6pt]
&&P^{(+)}_{12}T_2(u)\,T_1(u-\eta)P^{(+)}_{12}=\prod_{l=1}^{N}(u-\theta_l) \, T^{(1,\frac{1}{2})}_{\{12\}}(u),\label{Fused-Mono}
\eea
where the functions $a(u)$ and $d(u)$ are given by (\ref{a-d-functions}), and the fused monodromy matrix $T^{(1,\frac{1}{2})}_{\{12\}}(u)$ can be expressed in terms of the fused $R$-matrix $R^{(1,\frac{1}{2})}_{\{12\}\,3}(u)$ given by (\ref{Fusion-R})
\bea
T^{(1,\frac{1}{2})}_{\{12\}}(u)=R^{(1,\frac{1}{2})}_{\{12\} \,N}(u-\theta_N)\cdots R^{(1,\frac{1}{2})}_{\{12\} \,1}(u-\theta_1).\label{Fused-Mono-1}
\eea
Let us take the product of the transfer matrix $t(u)$ and $t(u-\eta)$ given by (\ref{trans-per})
\bea
t(u)\,t(u-\eta)&=&tr_{12}\lt\{e^{\frac{h\b}{2}(\sigma_1^z+\sigma_2^z)}\,T_2(u)\,T_1(u-\eta)\rt\}\no\\[4pt]
&=&tr_{12}\lt\{e^{\frac{h\b}{2}(\sigma_1^z+\sigma_2^z)}\,T_2(u)\,T_1(u-\eta)(P^{(-)}_{12}+P^{(+)}_{12})\rt\}\no\\[4pt]
&=&tr_{12}\lt\{P^{(-)}_{12}e^{\frac{h\b}{2}(\sigma_1^z+\sigma_2^z)}\,T_2(u)\,T_1(u-\eta)P^{(-)}_{12}\rt\}\no\\[4pt]
&&\quad\quad+tr_{12}\lt\{P^{(+)}_{12}e^{\frac{h\b}{2}(\sigma_1^z+\sigma_2^z)}\,T_2(u)\,T_1(u-\eta)P^{(+)}_{12}\rt\}\no\\[4pt]
&\stackrel{(\ref{Q-Det-1-1})}{=}&a(u)\,d(u-\eta)\times {\rm id}\no\\[4pt]
&&\quad\quad+tr_{12}\lt\{P^{(+)}_{12}e^{\frac{h\b}{2}(\sigma_1^z+\sigma_2^z)}P^{(+)}_{12}\,P^{(+)}_{12}\,T_2(u)\,T_1(u-\eta)P^{(+)}_{12}\rt\}\no\\[4pt]
&\stackrel{(\ref{Fused-Mono})}{=}&a(u)\,d(u-\eta)\times {\rm id}+e^{\frac{h\b}{2}}d(u)\,tr_{12}\lt\{P^{(+)}_{12}e^{\frac{h\b}{2}\sigma_{\{12\}}^z}P^{(+)}_{12}
T^{(1,\frac{1}{2})}_{\{12\}}(u)\rt\}.
\label{Proof}
\eea
Then the fused transfer matrix $\mathbb{W}(u)$ is given by the tracing over the subspace $W^{(+)}$ of the product of the fused monodromy matrix $T^{(1,\frac{1}{2})}_{\{12\}}(u)$ and $e^{\frac{h\b}{2}\sigma_{\{12\}}^z}$
\bea
\mathbb{W}(u)=tr_{12}\lt\{e^{\frac{h\b}{2}\sigma_{\{12\}}^z}\,T^{(1,\frac{1}{2})}_{\{12\}}(u)\rt\}.\label{W-op}
\eea
From the construction (\ref{Fusion-R-1}) and (\ref{Fused-Mono-1}), we know that the matrix elements of the fused monodromy matrix $T^{(1,\frac{1}{2})}_{\{12\}}(u)$, as a function of $u$, are operator-valued  polynomials of degrees up to $N$. Finally we have completed the proof of the
$t-W$ relation (\ref{t-W-relation-op}).

Using the similar fusion procedure as that we have done for the  $SU(2)$ case, we can also derive the   associated $t-W$ relations  (\ref{t-W-su(n)}) among  the fused transfer   matrices $\{t_i^{(Q)}(u)|i=2,\cdot\cdot\cdot,n-1\}$  and their counterparts $\{\mathbb{W}_{i}^{(Q)}(u)|i=1,\cdot\cdot\cdot,n-1\}$.


\section*{Appendix B: Numerical scheme of the spin-$\frac{1}{2}$ XXX closed chain}
\setcounter{equation}{0}
\renewcommand{\theequation}{B.\arabic{equation}}

For the convenience, let us introduce the function $\xi(u)$
\bea
\xi(u)= \ln\lt((q(u)+(4\cosh^2\frac{h\b}{2}-1)e^{-\b \bar{\e}(u)})/4\cosh^2\frac{h\b }{2}\rt).
\eea
Let us introduce two small positive parameters $\delta$ and $\Delta$ such that $0<\delta<\Delta<\frac{1}{2}$, and we can deform the integral contours in (\ref{Integral-Main}) without changing values of the resulting integrals as follows.
The decomposition (the first identity of (\ref{t-W-relation-Eig})) implies that the function $\xi(u)$ has singularities only on the straight lines ${\rm Im}(u)=\pm\frac{1}{2},\,\pm\frac{3}{2}$ and vanishes asymptotically, i.e., $\lim_{u\rightarrow\infty}\xi(u)=0$,  which allows us
to deform the integral contour ${\mathcal{C}_1}$ along the straight line ${\rm Im}(v)=(\frac{1}{2}-\delta)$ and the line ${\rm Im}(v)=(\frac{1}{2}+\delta)$ (
the contour ${\mathcal{C}_2}$ along the straight line ${\rm Im}(v)=-(\frac{1}{2}+\delta)$ and the line ${\rm Im}(v)=-(\frac{1}{2}-\delta)$ anti-clockwise without changing the integral values in (\ref{Integral-Main}). Namely, we can have the integral representation
\bea
\hspace{-2.2truecm}&&\hspace{-2.2truecm}\ln(q(u) \hspace{-0.02truecm}+\hspace{-0.02truecm}(4\cosh^2\frac{h\b}{2}-1)e^{-\b \bar{\e}(u)})=2\ln2\cosh\frac{h\b}{2}\no\\[8pt]
&&\quad\quad-\frac{1}{2\pi i}\int^{+\infty}_{-\infty}\,dv (\frac{1}{u-v-(\frac{1}{2}+\delta)\eta}+\frac{1}{u-v-(\frac{3}{2}+\delta)\eta})\xi(v+(\frac{1}{2}+\delta)\eta)\no\\[8pt]
&&\quad\quad+\frac{1}{2\pi i}\int^{+\infty}_{-\infty}\,dv (\frac{1}{u-v-(\frac{1}{2}-\delta)\eta}+\frac{1}{u-v-(\frac{3}{2}-\delta)\eta})\xi(v+(\frac{1}{2}-\delta)\eta)\no\\[8pt]
&&\quad\quad-\frac{1}{2\pi i}\int^{+\infty}_{-\infty}\,dv (\frac{1}{u-v+(\frac{1}{2}-\delta)\eta}+\frac{1}{u-v+(\frac{3}{2}-\delta)\eta})\xi(v-(\frac{1}{2}-\delta)\eta)\no\\[8pt]
&&\quad\quad+\frac{1}{2\pi i}\int^{+\infty}_{-\infty}\,dv (\frac{1}{u-v+(\frac{1}{2}+\delta)\eta}+\frac{1}{u-v+(\frac{3}{2}+\delta)\eta})\xi(v-(\frac{1}{2}+\delta)\eta).\label{Integral-rep-2}
\eea
The above new integral representation allows us to compute the values of $\bar{\e}(u\pm(\frac{1}{2}+\Delta)\eta)$ with $u\in \mathbb{R}$ provided that the values of $\xi(u)$ on the
four straight lines ${\rm Im}(v)=\pm(\frac{1}{2}-\delta),\,\pm(\frac{1}{2}+\delta)$ are known.   With the help of the analytical property (\ref{Property-epsion-function-1}) of the function $\bar{\e}(u)$ and the Cauchy's theorem, we can compute the values $\bar{\e}(u)$ on the four straight lines ${\rm Im}(u)=\pm(\frac{1}{2}-\delta),\,\pm(\frac{1}{2}+\delta)$ if we know its values on  the two straight lines ${\rm Im}(v)=\pm(\frac{1}{2}+\Delta)$. Namely, we have

\begin{footnotesize}
\bea
\bar{\e}(u+(\frac{1}{2}+\delta)\eta)\hspace{-0.2truecm}&=&\hspace{-0.2truecm}\frac{1}{2\pi i}\int^{+\infty}_{-\infty}dv\, \frac{\bar{\e}(v+(\frac{1}{2}+\Delta)\eta)}{u-v-(\Delta-\delta)\eta}
-\frac{1}{2\pi i}\int^{+\infty}_{-\infty}dv\, \frac{\bar{\e}(v-(\frac{1}{2}+\Delta)\eta)}{u-v+(1+\Delta+\delta)\eta},\quad u\in\mathbb{R},\label{Cauthy-1}\\[8pt]
\bar{\e}(u+(\frac{1}{2}-\delta)\eta)\hspace{-0.2truecm}&=&\hspace{-0.2truecm}\frac{1}{2\pi i}\int^{+\infty}_{-\infty}dv\, \frac{\bar{\e}(v+(\frac{1}{2}+\Delta)\eta)}{u-v-(\Delta+\delta)\eta}
-\frac{1}{2\pi i}\int^{+\infty}_{-\infty}dv\, \frac{\bar{\e}(v-(\frac{1}{2}+\Delta)\eta)}{u-v+(1+\Delta-\delta)\eta},\quad u\in\mathbb{R},\\[8pt]
\bar{\e}(u-(\frac{1}{2}-\delta)\eta)\hspace{-0.2truecm}&=&\hspace{-0.2truecm}\frac{1}{2\pi i}\int^{+\infty}_{-\infty}dv\, \frac{\bar{\e}(v+(\frac{1}{2}+\Delta)\eta)}{u-v-(1+\Delta-\delta)\eta}
-\frac{1}{2\pi i}\int^{+\infty}_{-\infty}dv\, \frac{\bar{\e}(v-(\frac{1}{2}+\Delta)\eta)}{u-v+(\Delta+\delta)\eta},\quad u\in\mathbb{R},\\[8pt]
\bar{\e}(u-(\frac{1}{2}+\delta)\eta)\hspace{-0.2truecm}&=&\hspace{-0.2truecm}\frac{1}{2\pi i}\int^{+\infty}_{-\infty}dv\, \frac{\bar{\e}(v+(\frac{1}{2}+\Delta)\eta)}{u-v-(1+\Delta+\delta)\eta}
-\frac{1}{2\pi i}\int^{+\infty}_{-\infty}dv\, \frac{\bar{\e}(v-(\frac{1}{2}+\Delta)\eta)}{u-v+(\Delta-\delta)\eta},\quad u\in\mathbb{R}.\label{Cauthy-2}
\eea
\end{footnotesize}%
Now our numerical  strategy can be constructed as follows. Starting from $\bar{\e}_{(n)}(u\pm(\frac{1}{2}+\Delta)\eta)$, we can compute the values $\bar{\e}_{(n)}(u\pm(\frac{1}{2}-\delta)\eta)$ and $\bar{\e}_{(n)}(u\pm(\frac{1}{2}+\delta)\eta)$ with the help of the Cauchy's integrals (\ref{Cauthy-1})-(\ref{Cauthy-2}). The integral representation (\ref{Integral-rep-2}) allows us to obtain $\bar{\e}_{(n+1)}(u\pm(\frac{1}{2}+\Delta)\eta)$. Then repeat the above step again. Finally we might reach the solution of the integral equation (\ref{Integral-Main}) with the analytic property (\ref{Property-epsion-function-1}).


\section*{Appendix C: High-temperature expansion of the Heisenberg chain}
\setcounter{equation}{0}
\renewcommand{\theequation}{C.\arabic{equation}}
For $\beta\rightarrow0$, the function $\bar{w}(u)$ becomes independent of $u$ since the integrand in relation (\ref{Integral-Main}) has no poles in the area surrounded by the contours $\mathcal{C}_1$ and $\mathcal{C}_2$. Inserting $\bar{\epsilon}(u)\sim 0$ into the integral in relation (\ref{free energy}) leads to the correct high-temperature entropy $-\beta f=\ln \Lambda(0)\sim\ln2$.

For small values of $\beta$, we seek $\bar{\epsilon}(u)$ as the series expansion
\bea
\bar{\e}(u)=\bar{\e}_1(u)+\beta\bar{\e}_2(u)+\cdot\cdot\cdot .
\eea
With regard to the expansion formula
\bea
\ln[q(u)+ae^{-\beta\bar{\e}}]=\hspace{-0.6truecm}&&\ln(a+1)+\frac{\beta}{a+1}[\frac{2J}{u^2+\frac{1}{4}}-a\bar{\e}_1]\no\\[4pt]
&&+\frac{\beta^2}{2(a+1)^2}[a(\frac{2J}{u^2+\frac{1}{4}}+\bar{\e}_1)^2-2(a^2+a)\bar{\e}_2]+\cdot\cdot\cdot,
\eea
where we have set $a=4\cosh^2(\frac{h\beta}{2})-1$ and the integral equation (\ref{Integral-Main}) transforms itself into an infinite sequence of coupled equations for the expansion functions \{$\bar{\e}_j(u)$\}:
\bea
\hspace{-2.8truecm}a\bar{\e}_1(u)=\frac{2J}{u^2+\frac{1}{4}}\hspace{-0.6truecm}&&-\frac{1}{2\pi i}\oint _{\mathcal{C}_1}dv (\frac{1}{u-v}
\hspace{-0.02truecm}+\frac{1}{u-v-i})
\hspace{-0.02truecm}\lt(\frac{2J}{v^2+\frac{1}{4}}-a\bar{\e}_1(v)\rt)\no\\[8pt]
\hspace{-1.2truecm}\hspace{-0.6truecm}&&+\frac{1}{2\pi i}\oint _{\mathcal{C}_2}dv (\frac{1}{u-v+i}
\hspace{-0.02truecm}+\frac{1}{u-v})
\hspace{-0.02truecm}\lt(\frac{2J}{v^2+\frac{1}{4}}-a\bar{\e}_1(v)\rt),\label{epsilon1}
\eea
\bea
\hspace{-0.3truecm}2(a+1)\bar{\e}_2(u)=\hspace{-0.6truecm}&&(\frac{2J}{u^2+\frac{1}{4}}-\bar{\e}_1(u))^2\no\\
\hspace{-0.6truecm}&&-\frac{1}{2\pi i}\oint _{\mathcal{C}_1}dv (\frac{1}{u-v}
+\frac{1}{u-v-i})
\lt((\frac{2J}{v^2+\frac{1}{4}}-\bar{\e}_1(v))^2-2(a+1)\bar{\e}_2(v)\rt)\no\\[8pt]
\hspace{-0.6truecm}&&+\frac{1}{2\pi i}\oint _{\mathcal{C}_2}dv (\frac{1}{u-v+i}
+\frac{1}{u-v})
\lt((\frac{2J}{v^2+\frac{1}{4}}-\bar{\e}_1(v))^2-2(a+1)\bar{\e}_2(v)\rt).\label{epsilon2}
\eea
etc. Note that the contour integrals of \{$\bar{\e}_j(u)$\} in RHS vanishes because the function $\bar{\e}(u)$ is analytic except some singularities on the axis ${\rm Im}(u)=\pm\frac{3}{2}$. Therefore, Eq.\,(\ref{epsilon1}) have two poles of $q(u)$ inside the contour $\mathcal{C}_1$ and  $\mathcal{C}_2$ respectively. Using the residue theorem, we obtain the expression of $\bar{\e}_1(u)$
\bea
\bar{\e}_1(u)=-\frac{1}{a}\frac{6J}{u^2+\frac{9}{4}}.
\eea
Substituting $\bar{\e}_1(u)$ into Eq.\,(\ref{epsilon2}) , we have
\bea
\bar{\e}_2(u)=-\frac{1}{2(a^2+a)}\lt(9(a-\frac{1}{a})(\frac{2J}{u^2+\frac{9}{4}})^2+4(a-3)\frac{(2J)^2}{u^2+\frac{9}{4}}\rt).
\eea
The above expressions allow us to obtain the HTE (\ref{HTE-XXX}) of the free energy.


\section*{Appendix D: Bethe ansatz solution of the $SU(3)$ model}
\setcounter{equation}{0}
\renewcommand{\theequation}{D.\arabic{equation}}

For the  $SU(3)$ model, it is well-known that eigenvalue of the quantum transfer matrix  $t_1^{(Q)}(u)$ given by (\ref{t-Q1-su3}) can be obtained by the algebraic Bethe ansatz method \cite{kor97}, where $\Lambda_1^{(Q)}(u)$ is given in terms of a homogeneous $T-Q$ relation, namely,
\bea
\Lambda_1(u)=\hspace{-0.6truecm}&&[(u-\eta\tau)(u+\eta\tau)]^{\frac{N}{2}}\frac{Q_1(u+\eta)Q_2(u-\eta)}{Q_1(u)Q_2(u)}\nonumber\\
\hspace{-0.6truecm}&&+[(u-\eta\tau)(u+\eta\tau-\eta)]^{\frac{N}{2}}\frac{Q_2(u+\eta)}{Q_2(u)}\nonumber\\
\hspace{-0.6truecm}&&+[(u-\eta\tau+\eta)(u+\eta\tau)]^{\frac{N}{2}}\frac{Q_1(u-\eta)}{Q_1(u)},\label{T-Q-su31}\\[4pt]
Q_1(u)=\prod_{j=1}^{L_1}\hspace{-0.6truecm}&&(u-\l_j^{(1)}), \quad Q_2(u)=\prod_{j=1}^{L_2}(u-\l_j^{(2)})\quad L_1,L_2=0,\cdots, N.\no
\eea
The two sets of parameters $\{\l_j^{(1)}|j=1,\cdots,L_1; L_1=0,\cdots,N\}$ and $\{\l_j^{(2)}|j=1,\cdots,L_2; L_2=0,\cdots,N\}$ satisfy the BAEs
\bea
\Big(\frac{\lambda_i^{(1)}-\eta\tau+\eta}{\lambda_i^{(1)}-\eta\tau}\Big)^{\frac{N}{2}}=-\prod\limits_{j=1}^{L_2}\frac{\lambda_i^{(1)}-\lambda_j^{(2)}-\eta}{\lambda_i^{(1)}-\lambda_j^{(2)}}\prod\limits_{k=1}^{L_1}\frac{\lambda_i^{(1)}-\lambda_k^{(1)}+\eta}{\lambda_i^{(1)}-\lambda_k^{(1)}-\eta},\quad i=1,...,L_1,\\
\Big(\frac{\lambda_l^{(2)}+\eta\tau-\eta}{\lambda_l^{(1)}+\eta\tau}\Big)^{\frac{N}{2}}=-\prod\limits_{k=1}^{L_1}\frac{\lambda_l^{(2)}-\lambda_k^{(1)}+\eta}{\lambda_l^{(2)}-\lambda_k^{(1)}}\prod\limits_{j=1}^{L_2}\frac{\lambda_l^{(2)}-\lambda_j^{(2)}-\eta}{\lambda_l^{(2)}-\lambda_j^{(2)}+\eta},\quad l=1,...,L_2.
\eea
It was shown \cite{Ess05} that the  eigenvalue $\Lambda_1^{(Q)}(u)$ corresponding to the state with $|\Lambda_1^{(Q)}(0)|_{max}$ belongs to the sector of $L_1=L_2=\frac{N}{2}$ with all Bethe roots having the small imaginary part and $\l_j^{(2)}=(\l_j^{(1)})^*$. Moreover ${\rm Im}(\l_j^{(i)})\sim 0$ for a large $N$, the $T-Q$ relation (\ref{T-Q-su31}) allows us to express $\bar{\Lambda}_1^{(Q)}(u)$  as
\bea
\bar{\Lambda}_1^{(Q)}(u)=\hspace{-0.6truecm}&&\frac{[(u-\eta\tau)(u+\eta\tau)]^{\frac{N}{2}}}{\prod_{j=1}^{\frac{N}{2}}(u-\lambda_j^{(1)})(u-\lambda_j^{(2)})} \frac{\prod_{j=1}^{\frac{N}{2}}(u-\lambda_j^{(2)}-\eta)(u-\lambda_j^{(1)}+\eta)}{[(u+\tau-\eta)(u-\tau+\eta)]^{\frac{N}{2}}}\nonumber\\
\hspace{-0.6truecm}&&+\frac{(u-\eta\tau)^{\frac{N}{2}}}{\prod_{j=1}^{\frac{N}{2}}(u-\lambda_j^{(2)})} \frac{\prod_{j=1}^{\frac{N}{2}}(u-\lambda_j^{(2)}+\eta)}{(u-\eta\tau+\eta)^{\frac{N}{2}}}\nonumber\\
\hspace{-0.6truecm}&&+\frac{(u+\eta\tau)^{\frac{N}{2}}}{\prod_{j=1}^{\frac{N}{2}}(u-\lambda_j^{(1)})} \frac{\prod_{j=1}^{\frac{N}{2}}(u-\lambda_j^{(1)}-\eta)}{(u+\eta\tau-\eta)^{\frac{N}{2}}},\label{T-Q-su3-1}
\eea
For the fused quantum transfer matrix $t_2^{(Q)}(u)$,  the corresponding eigenvalue $\Lambda_2^{(Q)}(u)$ can be expressed in terms of the $T-Q$ relation
\bea
\Lambda_2(u)=\hspace{-0.6truecm}&&[(u-\eta\tau+\frac{3}{2}\eta)(u+\eta\tau-\frac{3}{2}\eta)]^{\frac{N}{2}}\frac{Q_1(u-\frac{\eta}{2})Q_2(u+\frac{\eta}{2})}{Q_1(u+\frac{\eta}{2})Q_2(u-\frac{\eta}{2})}\nonumber\\
\hspace{-0.6truecm}&&+[(u-\eta\tau+\frac{3}{2}\eta)(u+\eta\tau-\frac{\eta}{2})]^{\frac{N}{2}}\frac{Q_2(u-\frac{3}{2}\eta))}{Q_2(u-\frac{\eta}{2})}\nonumber\\
\hspace{-0.6truecm}&&+[(u-\eta\tau+\frac{\eta}{2})(u+\eta\tau-\frac{3}{2}\eta)]^{\frac{N}{2}}\frac{Q_1(u+\frac{3}{2}\eta)}{Q_1(u+\frac{\eta}{2})},\label{T-Q-su32}
\eea
and the relation also allows us to express $\bar{\Lambda}_2^{(Q)}(u)$ as
\bea
\bar{\Lambda}_2^{(Q)}(u)=\hspace{-0.6truecm}&&\prod_{j=1}^{\frac{N}{2}}\frac{(u-\lambda_j^{(2)}+\frac{\eta}{2})(u-\lambda_j^{(1)}-\frac{\eta}{2})}{(u-\lambda_j^{(2)}-\frac{\eta}{2})(u-\lambda_j^{(1)}+\frac{\eta}{2})}\nonumber\\
\hspace{-0.6truecm}&&+\frac{(u+\eta\tau-\frac{\eta}{2})^{\frac{N}{2}}}{\prod_{j=1}^{\frac{N}{2}}(u-\lambda_j^{(2)}-\frac{\eta}{2})} \frac{\prod_{j=1}^{\frac{N}{2}}(u-\lambda_j^{(2)}-\frac{3}{2}\eta)}{(u+\eta\tau-\frac{3}{2}\eta)^{\frac{N}{2}}}\nonumber\\
\hspace{-0.6truecm}&&+\frac{(u-\eta\tau+\frac{\eta}{2})^{\frac{N}{2}}}{\prod_{j=1}^{\frac{N}{2}}(u-\lambda_j^{(1)}+\frac{\eta}{2})} \frac{\prod_{j=1}^{\frac{N}{2}}(u-\lambda_j^{(1)}+\frac{3}{2}\eta)}{(u-\eta\tau+\frac{3}{2}\eta)^{\frac{N}{2}}},\label{T-Q-su3-2}
\eea


\section*{Appendix E: Numerical  scheme of the $SU(3)$ model}
\setcounter{equation}{0}
\renewcommand{\theequation}{E.\arabic{equation}}

For the convenience, let us introduce the function $\xi_1(u)$ and $\xi_2(u)$
\bea
\xi_1(u)\hspace{-0.6truecm}&&=\ln\lt((q(u)e^{-\b \bar{\varepsilon}_2(u)}+(b(\beta)-1)e^{-\b \bar{\varepsilon}_3(u)})/b(\beta)\rt),\\
\xi_2(u)\hspace{-0.6truecm}&&=\ln\lt((e^{-\b \bar{\varepsilon}_1(u)}+(b(\beta)-1)e^{-\b \bar{\varepsilon}_4(u)})/b(\beta)\rt).
\eea
Similar as that we have done for the case of the XXX chain in Appendix B, let us introduce two small positive parameters $\delta$ and $\Delta$ such that $0<\delta<\Delta<\frac{1}{2}$, and we can deform the integral contours in (\ref{Integral-Main-su31})-(\ref{Integral-Main-su32}) without changing the values of resulting integrals as follows.
The decomposition (the first identity of (\ref{t-W-relation-su3-1})) implies that the function $\xi_1(u)$ has singularities only on the straight lines ${\rm Im}(u)=\pm\frac{1}{2},\,\pm\frac{3}{2}$ and vanishes asymptotically, i.e., $\lim_{u\rightarrow\infty}\xi_1(u)=0$, which allow us
to deform the integral contour ${\mathcal{C}_1^{\prime}}$ along the straight line ${\rm Im}(v)=(\frac{1}{2}-\delta)$ and the line ${\rm Im}(v)=(\frac{1}{2}+\delta)$ (the contour ${\mathcal{C}_2^{\prime}}$ along the straight line ${\rm Im}(v)=-(\frac{1}{2}+\delta)$ and the line ${\rm Im}(v)=-(\frac{1}{2}-\delta)$) anti-clockwise without changing the integral values in (\ref{Integral-Main-su31}). In the same way, we can also deform the integral contour ${\mathcal{C}_3^{\prime}}$ along the straight line ${\rm Im}(v)=(1-\delta)$ and the line ${\rm Im}(v)=(1+\delta)$ (the contour ${\mathcal{C}_4^{\prime}}$ along the straight line ${\rm Im}(v)=-(1+\delta)$ and the line ${\rm Im}(v)=-(1-\delta)$) anti-clockwise without changing the integral values in (\ref{Integral-Main-su32}).
Namely, we have four integral representations as follows:
\bea
\hspace{-2.2truecm}&&\hspace{-2.2truecm}\qquad\quad\ln(q(u)e^{-\b \bar{\varepsilon}_2(u)}+(b(\beta)-1)e^{-\b \bar{\varepsilon}_3(u)})=\ln b(\beta)\no\\[8pt]
&&\quad\quad-\frac{1}{2\pi i}\int^{+\infty}_{-\infty}\,dv (\frac{1}{u-v-(\frac{1}{2}+\delta)\eta}+\frac{1}{u-v-(\frac{3}{2}+\delta)\eta})\xi_1(v+(\frac{1}{2}+\delta)\eta)\no\\[8pt]
&&\quad\quad+\frac{1}{2\pi i}\int^{+\infty}_{-\infty}\,dv (\frac{1}{u-v-(\frac{1}{2}-\delta)\eta}+\frac{1}{u-v-(\frac{3}{2}-\delta)\eta})\xi_1(v+(\frac{1}{2}-\delta)\eta)\no\\[8pt]
&&\quad\quad-\frac{1}{2\pi i}\int^{+\infty}_{-\infty}\,dv (\frac{1}{u-v+(\frac{1}{2}-\delta)\eta}+\frac{1}{u-v+(\frac{3}{2}-\delta)\eta})\xi_1(v-(\frac{1}{2}-\delta)\eta)\no\\[8pt]
&&\quad\quad+\frac{1}{2\pi i}\int^{+\infty}_{-\infty}\,dv (\frac{1}{u-v+(\frac{1}{2}+\delta)\eta}+\frac{1}{u-v+(\frac{3}{2}+\delta)\eta})\xi_1(v-(\frac{1}{2}+\delta)\eta),\label{Integral-rep-su31-E}\\[8pt]
\bar{\varepsilon}_1(u)=\hspace{-0.6truecm}&&\frac{1}{2\pi i}\int^{+\infty}_{-\infty}\,dv  \lt(\frac{1}{u-v-(1+\delta)\eta}\xi_1(v+(\frac{1}{2}+\delta)\eta)-\frac{1}{u-v-(1-\delta)\eta})\xi_1(v+(\frac{1}{2}-\delta)\eta)\rt)\no\\[8pt]
\hspace{-0.6truecm}&&+\frac{1}{2\pi i}\int^{+\infty}_{-\infty}\,dv  \lt(\frac{1}{u-v+(1-\delta)\eta}\xi_1(v-(\frac{1}{2}-\delta)\eta)-\frac{1}{u-v+(1+\delta)\eta})\xi_1(v-(\frac{1}{2}+\delta)\eta)\rt),\label{Integral-rep-su32-E}\\[8pt]
\hspace{-2.2truecm}&&\hspace{-2.2truecm}\qquad\quad\ln(e^{-\b \bar{\varepsilon}_1(u)}+(b(\beta)-1)e^{-\b \bar{\varepsilon}_4(u)})=\ln b(\beta)\no\\[8pt]
&&\quad\quad-\frac{1}{2\pi i}\int^{+\infty}_{-\infty}\,dv (\frac{1}{u-v-(1+\delta)\eta}+\frac{1}{u-v-(2+\delta)\eta})\xi_2(v+(1+\delta)\eta)\no\\[8pt]
&&\quad\quad+\frac{1}{2\pi i}\int^{+\infty}_{-\infty}\,dv (\frac{1}{u-v-(1-\delta)\eta}+\frac{1}{u-v-(2-\delta)\eta})\xi_2(v+(1-\delta)\eta)\no\\[8pt]
&&\quad\quad-\frac{1}{2\pi i}\int^{+\infty}_{-\infty}\,dv (\frac{1}{u-v+(1-\delta)\eta}+\frac{1}{u-v+(2-\delta)\eta})\xi_2(v-(1-\delta)\eta)\no\\[8pt]
&&\quad\quad+\frac{1}{2\pi i}\int^{+\infty}_{-\infty}\,dv (\frac{1}{u-v+(1+\delta)\eta}+\frac{1}{u-v+(2+\delta)\eta})\xi_2(v-(1+\delta)\eta),\label{Integral-rep-su33-E}\\[8pt]
\bar{\varepsilon}_2(u)=\hspace{-0.6truecm}&&\frac{1}{2\pi i}\int^{+\infty}_{-\infty}\,dv  (\frac{1}{u-v-(\frac{3}{2}+\delta)\eta}\xi_2(v+(1+\delta)\eta)-\frac{1}{u-v-(\frac{3}{2}-\delta)\eta})\xi_2(v+(1-\delta)\eta))\no\\[8pt]
\hspace{-0.6truecm}&&+\frac{1}{2\pi i}\int^{+\infty}_{-\infty}\,dv  (\frac{1}{u-v+(\frac{3}{2}-\delta)\eta}\xi_2(v-(1-\delta)\eta)-\frac{1}{u-v+(\frac{3}{2}+\delta)\eta})\xi_2(v-(1+\delta)\eta)).\label{Integral-rep-su34-E}
\eea
The above new integral representations allow us to compute the values of $\bar{\varepsilon}_{\{1,4\}}(u\pm(1+\Delta)\eta)$, $\bar{\varepsilon}_{\{2,3\}}(u\pm(\frac{1}{2}+\Delta)\eta)$, $\bar{\varepsilon}_1(u\pm(1-\delta)\eta)$ and $\bar{\varepsilon}_1(u\pm(1+\delta)\eta)$  with $u\in \mathbb{R}$ provided that the values of $\xi_1(u)$ on the
four straight lines ${\rm Im}(v)=\pm(\frac{1}{2}-\delta),\,\pm(\frac{1}{2}+\delta)$ and the values of $\xi_2(u)$ on the other
four straight lines ${\rm Im}(v)=\pm(1-\delta),\,\pm(1+\delta)$ are known.  With the help of the analytical properties (\ref{Property-epsion-function-su3}) of the function $\{\bar{\varepsilon}_i(u)|i=1,...,4\}$ and the Cauchy's theorem, we can compute the values $\bar{\varepsilon}_{\{2,3\}}(u)$ on the four straight lines ${\rm Im}(u)=\pm(\frac{1}{2}-\delta),\,\pm(\frac{1}{2}+\delta)$ and the values $\bar{\varepsilon}_4(u)$ on the other four straight lines ${\rm Im}(u)=\pm(1-\delta),\,\pm(1+\delta)$ if we know the values $\bar{\varepsilon}_{\{2,3\}}(u)$ on the two straight lines ${\rm Im}(v)=\pm(\frac{1}{2}+\Delta)$ and the values $\bar{\varepsilon}_4(u)$ on the two straight lines ${\rm Im}(v)=\pm(1+\Delta)$. Namely, we have

\begin{footnotesize}
\bea
&&\bar{\varepsilon}_{\{2,3\}}(u+(\frac{1}{2}+\delta)\eta)=\frac{1}{2\pi i}\int^{+\infty}_{-\infty}dv\, \frac{\bar{\varepsilon}_{\{2,3\}}(v+(\frac{1}{2}+\Delta)\eta)}{u-v-(\Delta-\delta)\eta}
-\frac{1}{2\pi i}\int^{+\infty}_{-\infty}dv\, \frac{\bar{\varepsilon}_{\{2,3\}}(v-(\frac{1}{2}+\Delta)\eta)}{u-v+(1+\Delta+\delta)\eta},\quad u\in\mathbb{R},\label{Cauthy-su31}\\[8pt]
&&\bar{\varepsilon}_{\{2,3\}}(u+(\frac{1}{2}-\delta)\eta)=\frac{1}{2\pi i}\int^{+\infty}_{-\infty}dv\, \frac{\bar{\varepsilon}_{\{2,3\}}(v+(\frac{1}{2}+\Delta)\eta)}{u-v-(\Delta+\delta)\eta}
-\frac{1}{2\pi i}\int^{+\infty}_{-\infty}dv\, \frac{\bar{\varepsilon}_{\{2,3\}}(v-(\frac{1}{2}+\Delta)\eta)}{u-v+(1+\Delta-\delta)\eta},\quad u\in\mathbb{R},\\[8pt]
&&\bar{\varepsilon}_{\{2,3\}}(u-(\frac{1}{2}-\delta)\eta)=\frac{1}{2\pi i}\int^{+\infty}_{-\infty}dv\, \frac{\bar{\varepsilon}_{\{2,3\}}(v+(\frac{1}{2}+\Delta)\eta)}{u-v-(1+\Delta-\delta)\eta}
-\frac{1}{2\pi i}\int^{+\infty}_{-\infty}dv\, \frac{\bar{\varepsilon}_{\{2,3\}}(v-(\frac{1}{2}+\Delta)\eta)}{u-v+(\Delta+\delta)\eta},\quad u\in\mathbb{R},\\[8pt]
&&\bar{\varepsilon}_{\{2,3\}}(u-(\frac{1}{2}+\delta)\eta)=\frac{1}{2\pi i}\int^{+\infty}_{-\infty}dv\, \frac{\bar{\varepsilon}_{\{2,3\}}(v+(\frac{1}{2}+\Delta)\eta)}{u-v-(1+\Delta+\delta)\eta}
-\frac{1}{2\pi i}\int^{+\infty}_{-\infty}dv\, \frac{\bar{\varepsilon}_{\{2,3\}}(v-(\frac{1}{2}+\Delta)\eta)}{u-v+(\Delta-\delta)\eta},\quad u\in\mathbb{R},\\[8pt]
&&\bar{\varepsilon}_4(u+(1+\delta)\eta)=\frac{1}{2\pi i}\int^{+\infty}_{-\infty}dv\, \frac{\bar{\varepsilon}_4(v+(1+\Delta)\eta)}{u-v-(\Delta-\delta)\eta}
-\frac{1}{2\pi i}\int^{+\infty}_{-\infty}dv\, \frac{\bar{\varepsilon}_4(v-(1+\Delta)\eta)}{u-v+(2+\Delta+\delta)\eta},\quad u\in\mathbb{R},\\[8pt]
&&\bar{\varepsilon}_4(u+(1-\delta)\eta)=\frac{1}{2\pi i}\int^{+\infty}_{-\infty}dv\, \frac{\bar{\varepsilon}_4(v+(1+\Delta)\eta)}{u-v-(\Delta+\delta)\eta}
-\frac{1}{2\pi i}\int^{+\infty}_{-\infty}dv\, \frac{\bar{\varepsilon}_4(v-(1+\Delta)\eta)}{u-v+(2+\Delta-\delta)\eta},\quad u\in\mathbb{R},\\[8pt]
&&\bar{\varepsilon}_4(u-(1-\delta)\eta)=\frac{1}{2\pi i}\int^{+\infty}_{-\infty}dv\, \frac{\bar{\varepsilon}_4(v+(1+\Delta)\eta)}{u-v-(2+\Delta-\delta)\eta}
-\frac{1}{2\pi i}\int^{+\infty}_{-\infty}dv\, \frac{\bar{\varepsilon}_4(v-(1+\Delta)\eta)}{u-v+(\Delta+\delta)\eta},\quad u\in\mathbb{R},\\[8pt]
&&\bar{\varepsilon}_4(u-(1+\delta)\eta)=\frac{1}{2\pi i}\int^{+\infty}_{-\infty}dv\, \frac{\bar{\varepsilon}_4(v+(1+\Delta)\eta)}{u-v-(2+\Delta+\delta)\eta}
-\frac{1}{2\pi i}\int^{+\infty}_{-\infty}dv\, \frac{\bar{\varepsilon}_4(v-(1+\Delta)\eta)}{u-v+(\Delta-\delta)\eta},\quad u\in\mathbb{R}.\label{Cauthy-su32}
\eea
\end{footnotesize}%
Now our numerical  strategy can be constructed as follows. Starting from $\bar{\varepsilon}^{(n)}_{\{1,4\}}(u\pm(1+\Delta)\eta)$, $\bar{\varepsilon}^{(n)}_{\{2,3\}}(u\pm(\frac{1}{2}+\Delta)\eta)$, $\bar{\varepsilon}^{(n)}_1(u\pm(1-\delta)\eta)$ and $\bar{\varepsilon}^{(n)}_1(u\pm(1+\delta)\eta)$, we can compute the values $\bar{\varepsilon}_{\{2,3\}}^{(n)}(u\pm(\frac{1}{2}-\delta)\eta)$, $\bar{\varepsilon}_{\{2,3\}}^{(n)}(u\pm(\frac{1}{2}+\delta)\eta)$, $\bar{\varepsilon}_4^{(n)}(u\pm(1-\delta)\eta)$ and $\bar{\varepsilon}_4^{(n)}(u\pm(1+\delta)\eta)$ with the help of the Cauchy's integrals (\ref{Cauthy-su31})-(\ref{Cauthy-su32}). The integral representation (\ref{Integral-rep-2}) allows us to obtain $\bar{\varepsilon}^{(n+1)}_{\{1,4\}}(u\pm(1+\Delta)\eta)$, $\bar{\varepsilon}^{(n+1)}_{\{2,3\}}(u\pm(\frac{1}{2}+\Delta)\eta)$, $\bar{\varepsilon}^{(n+1)}_1(u\pm(1-\delta)\eta)$ and $\bar{\varepsilon}^{(n+1)}_1(u\pm(1+\delta)\eta)$. Then repeat the above step again. Finally we might reach the solutions of the integral equations (\ref{Integral-Main-su31}) and (\ref{Integral-Main-su32}) with the analytic properties (\ref{Property-epsion-function-su3}).


\section*{Appendix F: High-temperature expansion of the SU(3) model}
\setcounter{equation}{0}
\renewcommand{\theequation}{F.\arabic{equation}}
For $\beta\rightarrow0$, the functions $\{\bar{\varepsilon}_i(u)|i=1,...,4\}$ become independent of $u$ since the integrand in relations (\ref{Integral-Main-su31}) and (\ref{Integral-Main-su32}) have no poles in the area surrounded by the contours $\{\mathcal{C}_i^{\prime}|i=1,...,4\}$. Inserting $\{\bar{\varepsilon}_i(u)\sim 0|i=1,...,4\}$ into the integral in relation (\ref{free-energy-su3}) leads to the correct high-temperature entropy $-\beta f=\ln \Lambda_1(0)\sim\ln3$.

For small values of $\beta$, we seek $\{\bar{\varepsilon}_i(u)|i=1,...,4\}$ as the series expansion
\bea
\bar{\varepsilon}_i(u)=\bar{\varepsilon}_{i1}(u)+\beta\bar{\varepsilon}_{i2}(u)+\cdot\cdot\cdot, \quad i=1,...,4.
\eea
With regard to the expansion formulas
\bea
\ln[q(u)e^{-\beta\bar{\varepsilon}_2}+(b-1)e^{-\beta\bar{\varepsilon}_3}]=\hspace{-0.6truecm}&&\ln b+\frac{\beta}{b}[\frac{2J}{u^2+\frac{1}{4}}-\bar{\varepsilon}_{21}-(b-1)\bar{\varepsilon}_{31}]\no\\
\hspace{-0.6truecm}&&+\frac{\beta^2}{2b^2}[(b-1)(\frac{2J}{u^2+\frac{1}{4}}-\bar{\varepsilon}_{21}+\bar{\varepsilon}_{31})^2-2b\bar{\varepsilon}_{22}-2(b^2-b)\bar{\varepsilon}_{32}]+\cdot\cdot\cdot,\\
\ln[e^{-\beta\bar{\varepsilon}_1}+(b-1)e^{-\beta\bar{\varepsilon}_4}]=\hspace{-0.6truecm}&&\ln b+\frac{\beta}{b}[-\bar{\varepsilon}_{11}-(b-1)\bar{\varepsilon}_{41}]\no\\
\hspace{-0.6truecm}&&+\frac{\beta^2}{2b^2}[(b-1)(\bar{\varepsilon}_{11}-\bar{\varepsilon}_{41})^2-2b\bar{\varepsilon}_{12}-2(b^2-b)\bar{\varepsilon}_{42}]+\cdot\cdot\cdot,
\eea
where we have set $b=2\cosh(\frac{h\beta}{2})+1$ and the integral equations (\ref{Integral-rep-su31}-\ref{Integral-Main-su32}) transforms theirself into an infinite sequence of coupled equations for the expansion functions \{$\bar{\varepsilon}_{ij}(u)|i=1,...,4$\}:
\bea
\hspace{-0.8truecm}b\bar{\varepsilon}_{11}(u)&=&-\frac{1}{2\pi i}\oint _{\mathcal{C}_1^{\prime}}dv \frac{1}{u-v-\frac{i}{2}} \lt(\frac{2J}{v^2+\frac{1}{4}}-\bar{\varepsilon}_{21}(v)-(b-1)\bar{\varepsilon}_{31}(v)\rt)\no\\[8pt]
\hspace{-1.2truecm}&&+\frac{1}{2\pi i}\oint _{\mathcal{C}_2^{\prime}}dv\frac{1}{u-v+\frac{i}{2}} \lt(\frac{2J}{v^2+\frac{1}{4}}-\bar{\varepsilon}_{21}(v)-(b-1)\bar{\varepsilon}_{31}(v)\rt),\label{epsilon-E1}\\[8pt]
\hspace{-0.8truecm}b\bar{\varepsilon}_{21}(u)&=&-\frac{1}{2\pi i}\oint _{\mathcal{C}_3^{\prime}}dv \frac{1}{u-v-\frac{i}{2}} \lt(-\bar{\varepsilon}_{11}(v)-(b-1)\bar{\varepsilon}_{41}(v)\rt)\no\\[8pt]
\hspace{-1.2truecm}&&+\frac{1}{2\pi i}\oint _{\mathcal{C}_4^{\prime}}dv\frac{1}{u-v+\frac{i}{2}} \lt(-\bar{\varepsilon}_{11}(v)-(b-1)\bar{\varepsilon}_{41}(v)\rt),\label{epsilon-E2}\\[8pt]
\hspace{-0.8truecm}(b-1)\bar{\varepsilon}_{31}(u)&=&\frac{2J}{u^2+\frac{1}{4}}-\bar{\varepsilon}_{21}(u)\no\\[8pt]
&&-\frac{1}{2\pi i}\oint _{\mathcal{C}_1^{\prime}}dv (\frac{1}{u-v}
\hspace{-0.02truecm}+\frac{1}{u-v-i})
\hspace{-0.02truecm}\lt(\frac{2J}{v^2+\frac{1}{4}}-\bar{\varepsilon}_{21}(v)-(b-1)\bar{\varepsilon}_{31}(v)\rt)\no\\[8pt]
\hspace{-1.2truecm}&&+\frac{1}{2\pi i}\oint _{\mathcal{C}_2^{\prime}}\hspace{-0.02truecm}dv (\frac{1}{u-v+i}
\hspace{-0.02truecm}+\frac{1}{u-v})
\hspace{-0.02truecm}\lt(\frac{2J}{v^2+\frac{1}{4}}-\bar{\varepsilon}_{21}(v)-(b-1)\bar{\varepsilon}_{31}(v)\rt),\label{epsilon-E3}\\[8pt]
\hspace{-0.8truecm}(b-1)\bar{\varepsilon}_{41}(u)&=&-\bar{\varepsilon}_{11}(u)\no\\[8pt]
&&-\frac{1}{2\pi i}\oint _{\mathcal{C}_3^{\prime}}dv (\frac{1}{u-v}
\hspace{-0.02truecm}+\frac{1}{u-v-i})
\hspace{-0.02truecm}\lt(-\bar{\varepsilon}_{11}(v)-(b-1)\bar{\varepsilon}_{41}(v)\rt)\no\\[8pt]
\hspace{-1.2truecm}&&+\frac{1}{2\pi i}\oint _{\mathcal{C}_4^{\prime}}\hspace{-0.02truecm}dv (\frac{1}{u-v+i}
\hspace{-0.02truecm}+\frac{1}{u-v})
\hspace{-0.02truecm}\lt(-\bar{\varepsilon}_{11}(v)-(b-1)\bar{\varepsilon}_{41}(v)\rt),\label{epsilon-E4}\\[8pt]
\hspace{-0.8truecm}2b^2\bar{\varepsilon}_{12}(u)&=&-\frac{1}{2\pi i}\oint _{\mathcal{C}_1^{\prime}}dv \frac{1}{u-v-\frac{i}{2}} \lt((\frac{2J}{v^2+\frac{1}{4}}-\bar{\varepsilon}_{21}(v)+\bar{\varepsilon}_{31}(v))^2-2b\bar{\varepsilon}_{22}(v)
-2(b^2-b)\bar{\varepsilon}_{32}(v)\rt)\no\\[8pt]
\hspace{-1.2truecm}&&+\frac{1}{2\pi i}\oint _{\mathcal{C}_2^{\prime}}dv\frac{1}{u-v+\frac{i}{2}} \lt((\frac{2J}{v^2+\frac{1}{4}}-\bar{\varepsilon}_{21}(v)+\bar{\varepsilon}_{31}(v))^2-2b\bar{\varepsilon}_{22}(v)-2(b^2-b)\bar{\varepsilon}_{32}(v)\rt),\label{epsilon-E5}\\[8pt]
\hspace{-3.8truecm}2b^2\bar{\varepsilon}_{22}(u)=\hspace{-0.6truecm}&&-\frac{1}{2\pi i}\oint _{\mathcal{C}_3^{\prime}}dv \frac{1}{u-v-\frac{i}{2}} \lt((\bar{\varepsilon}_{11}(v)-\bar{\varepsilon}_{41}(v))^2-2b\bar{\varepsilon}_{12}(v)-2(b^2-b)\bar{\varepsilon}_{42}(v)\rt)\no\\[8pt]
\hspace{-1.2truecm}&&+\frac{1}{2\pi i}\oint _{\mathcal{C}_4^{\prime}}dv\frac{1}{u-v+\frac{i}{2}} \lt((\bar{\varepsilon}_{11}(v)-\bar{\varepsilon}_{41}(v))^2-2b\bar{\varepsilon}_{12}(v)-2(b^2-b)\bar{\varepsilon}_{42}(v)\rt),\label{epsilon-E6}
\eea
\begin{small}
\bea
\hspace{-4truecm}2(b^2-b)\hspace{-0.62truecm}&&\bar{\varepsilon}_{32}(u)\hspace{-0.02truecm}=(b\hspace{-0.02truecm}-\hspace{-0.02truecm}1)(\frac{2J}{u^2+\frac{1}{4}}\hspace{-0.02truecm}-\hspace{-0.02truecm}\bar{\varepsilon}_{21}(u)\hspace{-0.02truecm}+\hspace{-0.02truecm}\bar{\varepsilon}_{31}(u))^2\hspace{-0.02truecm}-\hspace{-0.02truecm}2b\bar{\varepsilon}_{22}(u)\no\\[8pt]
&&\hspace{-0.02truecm}-\hspace{-0.02truecm}\frac{1}{2\pi i}\oint _{\mathcal{C}_1^{\prime}}dv (\frac{1}{u\hspace{-0.02truecm}-\hspace{-0.02truecm}v}
\hspace{-0.02truecm}+\hspace{-0.02truecm}\frac{1}{u\hspace{-0.02truecm}-\hspace{-0.02truecm}v\hspace{-0.02truecm}-\hspace{-0.02truecm}i})
\hspace{-0.02truecm}\lt((b-1)(\frac{2J}{v^2+\frac{1}{4}}\hspace{-0.02truecm}-\hspace{-0.02truecm}\bar{\varepsilon}_{21}(v)\hspace{-0.02truecm}+\hspace{-0.02truecm}\bar{\varepsilon}_{31}(v))^2\hspace{-0.02truecm}-\hspace{-0.02truecm}2b\bar{\varepsilon}_{22}(v)\hspace{-0.02truecm}-\hspace{-0.02truecm}2(b^2\hspace{-0.02truecm}-\hspace{-0.02truecm}b)\bar{\varepsilon}_{32}(v)\rt)\no\\[8pt]
\hspace{-1.2truecm}&&\hspace{-0.02truecm}+\hspace{-0.02truecm}\frac{1}{2\pi i}\oint _{\mathcal{C}_2^{\prime}}\hspace{-0.02truecm}dv (\frac{1}{u\hspace{-0.02truecm}-\hspace{-0.02truecm}v\hspace{-0.02truecm}+\hspace{-0.02truecm}i}
\hspace{-0.02truecm}+\hspace{-0.02truecm}\frac{1}{u\hspace{-0.02truecm}-\hspace{-0.02truecm}v})
\hspace{-0.02truecm}\lt((b-1)(\frac{2J}{v^2+\frac{1}{4}}\hspace{-0.02truecm}-\hspace{-0.02truecm}\bar{\varepsilon}_{21}(v)\hspace{-0.02truecm}+\hspace{-0.02truecm}\bar{\varepsilon}_{31}(v))^2\hspace{-0.02truecm}-\hspace{-0.02truecm}2b\bar{\varepsilon}_{22}(v)\hspace{-0.02truecm}-
\hspace{-0.02truecm}2(b^2\hspace{-0.02truecm}-\hspace{-0.02truecm}b)\bar{\varepsilon}_{32}(v)\rt),\label{epsilon-E7}\\[8pt]
\hspace{-3.8truecm}2(b^2-b)\hspace{-0.62truecm}&&\bar{\varepsilon}_{42}(u)\hspace{-0.02truecm}=(b\hspace{-0.02truecm}-\hspace{-0.02truecm}1)(\hspace{-0.02truecm}\bar{\varepsilon}_{11}(u)\hspace{-0.02truecm}-\hspace{-0.02truecm}\bar{\varepsilon}_{41}(u))^2\hspace{-0.02truecm}-\hspace{-0.02truecm}2b\bar{\varepsilon}_{12}(u)\no\\[8pt]
&&\hspace{-0.02truecm}-\hspace{-0.02truecm}\frac{1}{2\pi i}\oint _{\mathcal{C}_3^{\prime}}dv (\frac{1}{u\hspace{-0.02truecm}-\hspace{-0.02truecm}v}
\hspace{-0.02truecm}+\hspace{-0.02truecm}\frac{1}{u\hspace{-0.02truecm}-\hspace{-0.02truecm}v\hspace{-0.02truecm}-\hspace{-0.02truecm}i})
\hspace{-0.02truecm}\lt((b-1)(\hspace{-0.02truecm}\bar{\varepsilon}_{11}(v)\hspace{-0.02truecm}-\hspace{-0.02truecm}\bar{\varepsilon}_{41}(v))^2\hspace{-0.02truecm}-\hspace{-0.02truecm}2b\bar{\varepsilon}_{12}(v)\hspace{-0.02truecm}-\hspace{-0.02truecm}2(b^2\hspace{-0.02truecm}-\hspace{-0.02truecm}b)\bar{\varepsilon}_{42}(v)\rt)\no\\[8pt]
\hspace{-1.2truecm}&&\hspace{-0.02truecm}+\hspace{-0.02truecm}\frac{1}{2\pi i}\oint _{\mathcal{C}_4^{\prime}}\hspace{-0.02truecm}dv (\frac{1}{u\hspace{-0.02truecm}-\hspace{-0.02truecm}v\hspace{-0.02truecm}+\hspace{-0.02truecm}i}
\hspace{-0.02truecm}+\hspace{-0.02truecm}\frac{1}{u\hspace{-0.02truecm}-\hspace{-0.02truecm}v})
\hspace{-0.02truecm}\lt((b-1)(\hspace{-0.02truecm}\bar{\varepsilon}_{11}(v)\hspace{-0.02truecm}
-\hspace{-0.02truecm}\bar{\varepsilon}_{41}(v))^2\hspace{-0.02truecm}-\hspace{-0.02truecm}
2b\bar{\varepsilon}_{12}(v)\hspace{-0.02truecm}-\hspace{-0.02truecm}2(b^2\hspace{-0.02truecm}-\hspace{-0.02truecm}b)
\bar{\varepsilon}_{42}(v)\rt).\label{epsilon-E8}
\eea
\end{small}
Note that the contour integrals of $\{\bar{\varepsilon}_{ij}(u)|i=2,...,4\}$ in RHS vanishes because the analytic properties (\ref{Property-epsion-function-su3}) of the functions $\{\bar{\varepsilon}_i(u)|i=2,...,4\}$. Therefore, Eq.\,(\ref{epsilon-E1}) have two poles of $q(u)$ inside the contour $\mathcal{C}_1^{\prime}$ and  $\mathcal{C}_2^{\prime}$ respectively. Using the residue theorem, we obtain the expression of $\bar{\varepsilon}_{11}(u)$
\bea
\bar{\varepsilon}_{11}(u)=-\frac{1}{b}\frac{4J}{u^2+1}.
\eea
Substituting $\bar{\varepsilon}_{11}(u)$ into Eq.\,(\ref{epsilon-E2}) and applying once again the residue theorem, we have
\bea
\bar{\varepsilon}_{21}(u)=-\frac{1}{b}\frac{6J}{u^2+\frac{9}{4}}.
\eea
Through the similar processes, we can obtain the rest of the expressions $\{\bar{\varepsilon}_{ij}(u)|i=2,...,4\}$
\bea
&&\hspace{-0.6truecm}\bar{\varepsilon}_{31}(u)=-\frac{1}{b}\frac{6J}{u^2+\frac{9}{4}},\\[6pt]
&&\hspace{-0.6truecm}\bar{\varepsilon}_{41}(u)=-\frac{1}{b(b-1)}\frac{8J}{u^2+4},\\[6pt]
&&\hspace{-0.6truecm}\bar{\varepsilon}_{12}(u)=-\frac{b-1}{2b^2}\lt(\frac{4}{(u^2+1)^2}+\frac{2}{u^2+1}\rt)(2J)^2,\\[6pt]
&&\hspace{-0.6truecm}\bar{\varepsilon}_{22}(u)=-\frac{1}{2b^4}\lt(\frac{9(b^2-1)}{(u^2+\frac{9}{4})^2}+\frac{(4b^2-3b-9)}{u^2+\frac{9}{4}}\rt)(2J)^2,\\[6pt]
&&\hspace{-0.6truecm}\bar{\varepsilon}_{32}(u)=-\frac{1}{2b^4}\lt(\frac{9(b^3-b-1)}{(u^2+\frac{9}{4})^2}+\frac{(4b^4-4b^3-4b^2+3b+9)}{(b-1)(u^2+\frac{9}{4})}\rt)(2J)^2,\\[6pt]
&&\hspace{-0.6truecm}\bar{\varepsilon}_{42}(u)\hspace{-0.02truecm}=\hspace{-0.02truecm}-\hspace{-0.02truecm}\frac{1}{2(b^3\hspace{-0.02truecm}-\hspace{-0.02truecm}b^2)}\lt(\frac{16(b^2\hspace{-0.02truecm}-\hspace{-0.02truecm}
b\hspace{-0.02truecm}-\hspace{-0.02truecm}1)}{(b\hspace{-0.02truecm}-\hspace{-0.02truecm}1)(u^2\hspace{-0.02truecm}+\hspace{-0.02truecm}4)^2}
\hspace{-0.02truecm}+\frac{2(3b^2\hspace{-0.02truecm}-\hspace{-0.02truecm}2b\hspace{-0.02truecm}-\hspace{-0.02truecm}9)}{b(u^2\hspace{-0.02truecm}
+\hspace{-0.02truecm}4)}\rt)(2J)^2.
\eea
Substituting the above results into the free energy Eq.\,(\ref{free-energy-su3}) and integral, we have the HTE of the free energy
\bea
f/T=-\ln(2\cosh(\frac{h}{2T})+1)+\frac{J}{T}\frac{4}{2\cosh(\frac{h}{2T})+1}-\frac{J^2}{T^2}\frac{24\cosh(\frac{h}{2T})}{(2\cosh(\frac{h}{2T})+1)^2}+\cdot\cdot\cdot.
\eea
Finally we have completed the proof of (\ref{HTE-su(3)}).


%

\end{document}